\documentclass[12pt, twocolumn]{openjournal}

\usepackage{xcolor}
\usepackage{textgreek}
\usepackage[utf8]{inputenc}
\usepackage[english]{babel}
\usepackage{hyperref}
\hypersetup{
    colorlinks=true,
    linkcolor=blue,
    filecolor=blue,      
    urlcolor=red,
    citecolor=blue,
}
\usepackage{color,colortbl}
\usepackage{tensind}
\tensordelimiter{?}
\DeclareGraphicsExtensions{.bmp,.png,.jpg,.pdf}
\usepackage{verbatim}
\usepackage[normalem]{ulem}
\usepackage{orcidlink}
\usepackage{soul}

\usepackage{amsmath}

\graphicspath{ {./figs/} }

\begin{document}
\title{VENUS: Strong Lens Mass Models and Einstein Radii of 60 Galaxy Clusters}

\author{\vspace{-1.5cm}Masamune~Oguri$^{1, 2}$\orcidlink{0000-0003-3484-399X}}
\author{Vasily Kokorev$^{3, 4}$\orcidlink{0000-0002-5588-9156}}
\author{Shun Nishida$^{2}$\orcidlink{0009-0005-6923-5590}}
\author{Yoshinobu Fudamoto$^{1}$\orcidlink{0000-0001-7440-8832}} 
\author{Seiji Fujimoto$^{5, 6}$\orcidlink{0000-0001-7201-5066}}
\author{\\Dan Coe$^{7, 8, 9}$\orcidlink{0000-0001-7410-7669}} 
\author{Johan Richard$^{10}$\orcidlink{0000-0001-5492-1049}} 
\author{Jose M. Diego$^{11}$\orcidlink{0000-0001-9065-3926}} 
\author{Fengwu Sun$^{12}$\orcidlink{0000-0002-4622-6617}} 
\author{Lukas J. Furtak$^{3, 4}$\orcidlink{0000-0001-6278-032X}} 
\author{\\Maru\v{s}a Brada\v{c}$^{13, 14}$\orcidlink{0000-0001-5984-0395}} 
\author{Larry D. Bradley$^{7}$\orcidlink{0000-0002-7908-9284}} 
\author{John Chisholm$^{3, 4}$\orcidlink{0000-0002-0302-2577}} 
\author{Christopher J. Conselice$^{15}$\orcidlink{0000-0003-1949-7638}} 
\author{\\Miroslava Dessauges-Zavadsky$^{16}$\orcidlink{0000-0003-0348-2917}} 
\author{Henry C. Ferguson$^{7}$\orcidlink{0000-0001-7113-2738}} 
\author{Steven L. Finkelstein$^{3, 4}$\orcidlink{0000-0001-8519-1130}} 
\author{\\Brenda L. Frye$^{17}$\orcidlink{0000-0003-1625-8009}} 
\author{Mauro Gonz\'{a}lez-Otero$^{18}$\orcidlink{0000-0002-4837-1615}} 
\author{Tiger Yu-Yang Hsiao$^{3, 4}$\orcidlink{0000-0003-4512-8705}} 
\author{Kotaro Kohno$^{19, 20}$\orcidlink{0000-0002-4052-2394}} 
\author{\\Ray A. Lucas$^{7}$\orcidlink{0000-0003-1581-7825}} 
\author{Jos\'{e} M. Palencia$^{5}$\orcidlink{0000-0003-0942-817X}} 
\author{Richard Pan$^{21}$\orcidlink{0000-0002-9651-5716}} 
\author{Massimo Ricotti$^{22}$\orcidlink{0000-0003-4223-7324}} 
\author{\\Gregor Rihtar\v{s}i\v{c}$^{13}$\orcidlink{0009-0009-4388-898X}} 
\author{Luke Robbins$^{21}$\orcidlink{0000-0002-6265-2675}} 
\author{Chris Willott$^{23}$\orcidlink{0000-0002-4201-7367}} 

\email{masamune.oguri@chiba-u.jp}
\affiliation{$^{1}$Center for Frontier Science, Chiba University, 1-33 Yayoi-cho, Inage-ku, Chiba 263-8522, Japan}
\affiliation{$^{2}$Department of Physics, Graduate School of Science, Chiba University, 1-33 Yayoi-Cho, Inage-Ku, Chiba 263-8522, Japan}
\affiliation{$^{3}$Department of Astronomy, The University of Texas at Austin, 2515 Speedway, Stop C1400, Austin, TX 78712, USA}
\affiliation{$^{4}$Cosmic Frontier Center, The University of Texas at Austin, Austin, TX 78712, USA}
\affiliation{$^{5}$David A. Dunlap Department of Astronomy and Astrophysics, University of Toronto, 50 St. George Street, Toronto, Ontario, M5S 3H4, Canada}
\affiliation{$^{6}$Dunlap Institute for Astronomy and Astrophysics, 50 St. George Street, Toronto, Ontario, M5S 3H4, Canada}
\affiliation{$^{7}$Space Telescope Science Institute (STScI), 3700 San Martin Drive, Baltimore, MD 21218, USA}
\affiliation{$^{8}$Center for Astrophysical Sciences, Department of Physics and Astronomy, The Johns Hopkins University, 3400 N Charles St. Baltimore, MD 21218, USA}
\affiliation{$^{9}$Association of Universities for Research in Astronomy (AURA), Inc.~for the European Space Agency (ESA)}
\affiliation{$^{10}$Universit\'e Claude Bernard Lyon 1, CRAL UMR5574, ENS de Lyon, CNRS, Villeurbanne, F-69622, France}
\affiliation{$^{11}$Instituto e F\'isica de Cantabria,(CSIC-UC), Avda. Los Castros s/n. 39005, Santander, Spain}
\affiliation{$^{12}$Department of Astronomy, School of Science, Westlake University, Hangzhou, Zhejiang 310030, P. R. China}
\affiliation{$^{13}$University of Ljubljana, Faculty of Mathematics and Physics, Jadranska ulica 19, SI-1000 Ljubljana, Slovenia}
\affiliation{$^{14}$Department of Physics and Astronomy, University of California Davis, 1 Shields Avenue, Davis, CA 95616, USA}
\affiliation{$^{15}$Jodrell Bank Centre for Astrophysics, University of Manchester, Oxford Road, Manchester M13 9PL, UK}
\affiliation{$^{16}$Department of Astronomy, University of Geneva, Chemin Pegasi 51, 1290 Versoix, Switzerland}
\affiliation{$^{17}$Department of Astronomy/Steward Observatory, University of Arizona, 933 N. Cherry Avenue, Tucson, AZ 85721, USA}
\affiliation{$^{18}$Instituto de Astrof\'isica de Andaluc\'ia--CSIC, Glorieta de la Astronom\'ia s/n, E--18008 Granada, Spain}
\affiliation{$^{19}$Institute of Astronomy, Graduate School of Science, The University of Tokyo, 2-21-1 Osawa, Mitaka, Tokyo, 181-0015 Japan}
\affiliation{$^{20}$Research Center for the Early Universe, Graduate School of Science, The University of Tokyo, 7-3-1 Hongo, Bunkyo-ku, Tokyo 113-0033, Japan}
\affiliation{$^{21}$Department of Physics \& Astronomy, Tufts University, Medford, MA 02155, USA}
\affiliation{$^{22}$Department of Astronomy, University of Maryland, College Park, 20742, USA}
\affiliation{$^{23}$NRC Herzberg, 5071 West Saanich Rd, Victoria, BC V9E 2E7, Canada}

\begin{abstract}
We present the strong lens mass modeling result of 60 galaxy clusters
targeted by Vast Exploration for Nascent, Unexplored Sources
(VENUS), the James Webb Space Telescope (JWST) multi-epoch Treasury program.
We obtain refined mass models for the 60 clusters with the {\sc
  glafic} software by fitting positions of $\sim 4500$ multiple images,
$\sim 1150$ of which are newly identified multiple images with help of
multi-band JWST images. Each cluster contains typically $\sim 50$
multiple images, thanks to the selection of target clusters
based on their high lensing strengths.  We validate our models with the
Jackknife method.  The median Einstein radius is $\sim 30''$ and the
distribution of the Einstein radii extends out to $\sim 60''$ for the
source redshift of 9, which confirms that their lensing strengths are
indeed among the highest known to date. For the source redshift of 9,
we identify  MACS~J0717.5+3745, MACS~J0600.1$-$2008, and PLCK~G287.0+32.9 as
clusters with the largest Einstein radii as well as the largest areas
of the high-magnification ($|\mu|>10$) regions. We find that the
observed Einstein radii of VENUS clusters are fully compatible with
the standard cold dark matter model. 
\end{abstract}


\maketitle

\section{Introduction}
\label{sec:intro}

A massive cluster of galaxies can produce more than one hundred
multiple images of galaxies behind it by the gravitational lensing
effect, which enable various studies that are difficult without
gravitational lensing
\citep[see][for reviews]{2011A&ARv..19...47K,2024SSRv..220...19N}.
For example, the magnification and multiple
images help identify high-redshift galaxies
\citep[e.g.,][]{2013ApJ...762...32C,2023Sci...380..416W,2024Natur.626..975A}
as well as resolve their internal structures
\citep[e.g.,][]{2021A&A...646A..57V,2023ApJ...949L..34H,2024Natur.632..513A,2024Natur.636..332M,2025ApJ...995L..74B}.
Time delays between multiple images, which can be measured for
variable sources such as quasars and supernovae, not only allow us to
measure the Hubble constant
\citep[e.g.,][]{2023Sci...380.1322K,2025ApJ...979...13P,2026A&A...708A.291S}
but also open up the possibility of reconstructing the light curve of
a source from single-epoch imaging
\citep[e.g.,][]{2022Natur.611..256C,2025arXiv251205180Z}.
By utilizing microlensing, one can even detect individual stars in
gravitationally lensed arcs to study their stellar population
\citep[e.g.,][]{2018NatAs...2..334K,2025NatAs...9..428F}. Accurate and
precise strong lens mass models are crucial for all these applications.

In addition to studying lensed sources, strong gravitational lensing
probes mass distributions of lensing clusters, which provides a useful
means of testing the standard structure formation model and exploring
the nature of dark matter. For instance, the abundance of arcs in
observations have been confronted with theoretical predictions based
on the $\Lambda$ cold dark matter ($\Lambda$CDM) paradigm
\citep[e.g.,][for a review]{2013SSRv..177...31M}. The
density profiles of massive clusters measured by strong and weak
gravitational lensing have also been directly confronted with the
$\Lambda$CDM predictions to find good agreements between them
\citep[e.g.,][]{2012MNRAS.420.3213O,2013ApJ...765...24N}. Measurements
of the very central density profiles with strong gravitational lensing
place tight constraints on the dark matter self-interaction cross-section
\citep[e.g.,][]{2022MNRAS.510...54A}. The abundance of galaxy-galaxy
strong lenses near cluster centers also contains useful cosmological
information \citep[e.g.,][]{2020Sci...369.1347M}. Furthermore, the
statistics of the large Einstein radii serve as another useful test of the
$\Lambda$CDM paradigm \citep[e.g.,][]{2008MNRAS.390.1647B,2009MNRAS.392..930O}.

A significant advance in studies of cluster strong lensing is achieved by
the Hubble Frontier Fields \citep[HFF;][]{2017ApJ...837...97L} for which deep
imaging of six massive galaxy clusters with the Hubble Space Telescope
(HST) was conducted. Cluster Lensing And Supernova survey with Hubble
\citep[CLASH;][]{2012ApJS..199...25P} and Reionization Lensing Cluster
Survey \citep[RELICS;][]{2019ApJ...884...85C} provide HST imaging data
of clusters that are shallower than HFF but for larger numbers of
clusters. Massive clusters continue to be important targets in the era
of James Webb Space Telescope (JWST), see e.g.,
\citet{2026ConPh..67..203Z}. In addition to the galaxy
cluster SMACS J0723.3$-$7327 being one of five targets of the JWST
Early Release Observations \citep{2022ApJ...936L..14P}, there are
several related JWST programs including the Canadian NIRISS Unbiased Cluster
Survey \citep[CANUCS;][]{2026ApJS..282....3S}, Ultradeep NIRSpec and
NIRCam Observations before the Epoch of Reionization
\citep[UNCOVER;][]{2024ApJ...974...92B},
All the Little Things \citep[ALT;][]{2024arXiv241001874N}, GLIMPSE
\citep{2025arXiv251107542A}, GLASS-JWST \citep{2025A&A...699A.225W},
Prime Extragalactic Areas for Reionization and Lensing Science
\cite[PEARLS;][]{2023AJ....165...13W}, Cosmic Spring
\citep{2022ApJ...940L...1W}, Medium-band Astrophysics with
the Grism of NIRCam In Frontier Fields
\citep[MAGNIF;][]{2025ApJ...987..186F}, and Strong LensIng and Cluster
Evolution \citep[SLICE;][]{2026ApJ..1001...60C}.

Vast Exploration for Nascent, Unexplored Sources (VENUS) is a new
JWST program in Cycle 4 (GO-6882, PIs: S. Fujimoto and D. Coe) that
conducts extensive NIRCam imaging observations to construct a sample
of 60 clusters with uniform NIRCam images in 10 filters
(S. Fujimoto et al. in preparation). In this 
paper, we present strong lens mass models of all 60 VENUS
clusters. The mass modeling is conducted with the {\sc glafic} software
\citep{2010PASJ...62.1017O,2021PASP..133g4504O}, which shows the best
performance in a mass modeling comparison project using simulated
clusters \citep{2017MNRAS.472.3177M}. We then compare and discuss
their lensing strengths quantified by areas of high-magnification
regions as well as the Einstein radii. We compare the observed
Einstein radii with theoretical predictions based on the $\Lambda$CDM
paradigm. In a companion paper (J.  F. V. Allingham et al., in
preparation), independent mass models of 57 VENUS clusters using a
different software and technique are presented.

This paper is organized as follows. We describe the data and the
method used for the analysis in Section~\ref{sec:method}. The result
of the mass modeling is given in Section~\ref{sec:model}. We derive
lensing strengths of individual clusters in
Section~\ref{sec:strength}. The comparison of observed Einstein radii
with the prediction of the $\Lambda$CDM model is given in
Section~\ref{sec:ein}. Finally, we conclude in Section~\ref{sec:concl}.
The strong lens mass modeling is performed assuming the matter density
of $\Omega_{\mathrm{m}}=0.3$ and the cosmological constant of
$\Omega_{\Lambda}=0.7$, although we note that the mass modeling result
essentially does not depend on the choice of the cosmological
parameters. On the other hand, the $\Lambda$CDM model prediction of
the largest Einstein radius in the Universe is made assuming the
best-fitting cosmological parameters from the latest cosmological data
set \citep{2025JCAP...11..062L}.

\section{Data and Method}\label{sec:method}

\subsection{VENUS Data}\label{sec:venus_data}

In this paper, we use NIRCam imaging data obtained by the VENUS
program for our analysis. The NIRCam observations consist of imaging
in ten bands including F090W, F115W, F150W, F200W, F210M, F277W,
F300M, F356W, F410M, and F444W, with exposure times of $0.35$--$0.57$
hours per filter. The exposure time is chosen to achieve a 5$\sigma$ point
source detection limit of $\sim$28 mag across all the NIRCam filters. 

The procedure to reduce the data is as follows. We reduce JWST
level-2 products from the Mikulski Archive for Space Telescopes with
the {\sc grizli} pipeline \citep{2023zndo...8370018B}, similarly to
the public DAWN JWST Archive \citep{2023ApJ...947...20V}.
Additional post-processing steps include additional  background, $1/f$
noise, and diffraction spike subtraction procedures
\citep[e.g.,][]{2025ApJ...983L..22K}. The final drizzled image has a
pixel size of $0\farcs03$. The JWST images are supplemented by
archival Hubble Space Telescope images of the same clusters, which are
taken from Gaia-aligned mosaics from the CHArGE archive
\citep{2022ApJS..263...38K} and are drizzled in the same manner as the
JWST data.

From the reduced images, object detection and photometry in
different bands are performed using the {\sc sep} software
\citep{2016JOSS....1...58B,1996A&AS..117..393B}. Photometric redshifts
of detected objects are derived with the {\sc eazy} software
\citep{2008ApJ...686.1503B} with the (aperture-corrected) aperture
photometry with the aperture size of $0\farcs2$.

We note that JWST NIRCam imaging is yet to be conducted for three
clusters, RXC J0949.8+1707, Abell 1835, and RXC J1347.5$-$1145, for
which we use existing HST imaging data for our mass modeling. We plan
to update our strong lens mass models after JWST images of these
clusters are taken.

\subsection{Identification of Multiple Images}

Since the VENUS program selects target clusters based on the
availability of strong lens mass models, for each cluster, multiple
images that are previously identified are mostly available.
Spectroscopic redshifts of these previously identified multiple images
are taken from the literature when available.  After constructing an 
initial mass model for each cluster following the strategy described in
Section~\ref{sec:modeling_method}, we search for new multiple images
taking advantage of the high-quality VENUS data. Specifically, we
select multiple images based on the combination of image morphologies,
photometric redshifts, and the consistency of image configurations
with the current mass model. After identifying secure multiple
images, we add these new multiple images to refine the mass model, and
again search for new multiple images. We repeat this procedure
iteratively to finalize the multiple image set used for mass modeling.
In total, we use $\sim 4500$ multiple images for our mass modeling of
60 clusters, $\sim 1150$ of which are newly identified in this paper.
Our selection of multiple images for the mass modeling analysis is
rather conservative such that we adopt only high-confidence systems
that are judged from photometric redshifts, morphologies, and the
consistency with mass modeling. References for multiples images and
spectroscopic redshifts are listed in
Appendix~\ref{ap:ref}.\footnote{We will make lists of multiple images
publicly available at the time when the VENUS Data Release 1 (DR1) is
released.} 

\subsection{Mass Modeling Method}\label{sec:modeling_method}

We conduct strong lens mass modeling using the {\sc glafic} software
\citep{2010PASJ...62.1017O,2021PASP..133g4504O}, which takes the
so-called parametric approach. In parametric mass modeling, a lens
potential of a cluster is described by a superposition of lens
potentials of a small number of lens components. Those lens components
typically include dark matter halos, member galaxies, and various
perturbations. The resulting precision and accuracy of mass modeling 
critically depend on which lens components are assumed and which
parameters of those lens components are fit to the data. Here we
describe our strategy to determine a set of lens components and
parameters used for mass modeling of each cluster.

We use the elliptical \citet[][hereafter NFW]{1997ApJ...490..493N}
density profile for halo components. Its radial density profile is
given by
\begin{equation}
  \rho(r)=\frac{\rho_{\mathrm{s}}}{(r/r_{\mathrm{s}})(1+r/r_{\mathrm{s}})^2},
  \label{eq:def_nfw}
\end{equation}
where $\rho_{\mathrm{s}}$ and $r_{\mathrm{s}}$ are the characteristic
density and the scale radius, respectively. The ellipticity is
introduced in the isodensity contours of its surface mass density. We use
an approximation proposed in \citet{2021PASP..133g4504O} for fast
calculations of the lensing properties of the elliptical NFW profile.
Each halo component is characterized by six parameters, including the
mass, the concentration parameter, the centroid position, the
ellipticity, and the position angle. The mass and concentration
parameter are defined using the virial overdensity
\citep{1997PThPh..97...49N}. 

First, a halo component is added at the position of the brightest
cluster galaxy to optimize model parameters. Additional halo components
may be added at around positions of bright cluster member galaxies,
only if adding additional halo components significantly improves the
fit. All the parameters of each halo component, including its
centroid, are allowed to float when a sufficient number of multiple
images are found around it. In contrast, if the halo component is
located relatively in the outskirts of the strong lensing region, its
centroid position may be fixed to the position of a bright cluster
member galaxy. In that situation, the concentration parameter $c$ may
also be fixed to $c=10$. When included as model parameters, the
fitting range of the concentration parameter is always restricted to
$1<c<40$. The ellipticity $e$, which is defined by one minus the ratio
of the lengths of minor and major axes of the isodensity contour, is
always restricted to $0<e<0.8$. 

Cluster member galaxies are modeled by pseudo-Jaffe ellipsoids
\citep{1983MNRAS.202..995J,2001astro.ph..2341K}. Its radial density
profile is given by
\begin{equation}
  \rho(r)=\frac{\rho_{0}}{(r^2+r_{\mathrm{core}}^2)(r^2+r_{\mathrm{trun}}^2)},
  \label{eq:def_jaffe}
\end{equation}
where $\rho_0$ is the characteristic density and $r_{\mathrm{core}}$
and $r_{\mathrm{trun}}$ are the core and break radii,
respectively. Since we always fix the core radius to
$r_{\mathrm{core}}=0$ when modeling member galaxies, each pseudo-Jaffe
ellipsoid is characterized by six parameters, including the velocity
dispersion, the centroid position, the ellipticity, the position
angle, and the break radius.  

In order to reduce the number of parameters, we adopt a scaling
relation for some of the parameters characterizing the pseudo-Jaffe
ellipsoids. We do so by assuming that velocity dispersions and break
radii of individual pseudo-Jaffe ellipsoids scale with luminosities
$L$ of member galaxies as
\begin{align}
  \frac{\sigma}{\sigma_*} &=\left(\frac{L}{L_*}\right)^{1/4},\\
  \frac{r_{\mathrm{trun}}}{r_{\mathrm{trun},*}} &=\left(\frac{L}{L_*}\right)^{\eta},
\end{align}
and treat $\sigma_*$, $r_{\mathrm{trun},*}$, and $\eta$ as
parameters. We usually take the fiducial luminosity $L_*$ as that of
the brightest cluster galaxy. Ellipticities and position angles of
individual member galaxies are fixed to observed values for their
light profiles. These parameters are restricted to the ranges
$100\,\mathrm{km/s}<\sigma_*<400\,\mathrm{km/s}$,
$1''<r_{\mathrm{trun},*}<100''$, and $0.2<\eta<1.5$. The parameter
$\eta$ may be fixed to $\eta=1$ if there are not many multiple
images. Cluster member galaxies are selected from the imaging data
using either the color-magnitude diagram or photometric redshifts. 

In some rare cases, a cluster member galaxy may be modeled separately
without assuming the scaling relation if that member galaxy is
critically important for accurately reproducing some multiple
images. Examples of these rare cases include a member galaxy that
produces the Einstein cross images of the supernova Refsdal
\citep{2023Sci...380.1322K} and a member galaxy that produces the
three supernova images in Abell~370
\citep{2022Natur.611..256C}. Furthermore, when it is needed to
reproduce central images accurately, the brightest cluster galaxy may
be modeled separately assuming the elliptical
\citet{1990ApJ...356..359H} profile, although we note that for most of
the clusters the bright cluster galaxies are simply modeled by the
pseudo-Jaffe ellipsoids adopting the scaling relation mentioned above.

In addition, we include multipole perturbations to improve the
fit. Its lens potential is generally given by
\begin{equation}
  \phi= -\frac{\epsilon}{m}r^n\cos
  \left[m\left(\theta-\theta_\epsilon-\frac{\pi}{2}\right)\right],
\end{equation}
where, for given $n$ and $m$, $\epsilon$ and $\theta_\epsilon$ are
parameters. The case $m=2$ and $n=2$ corresponds to the external
shear. Throughout the paper, we fix $n=2$, and consider $m=2$ and
higher ($m=3$, $4$, $\ldots$) multipole perturbations. As for halo
components, we add higher order terms only when it significantly
improves the fit. 

We optimize model parameters so as to reproduce positions of multiple
images. Source redshifts are fixed to the spectroscopic redshifts when
available, otherwise they are also treated as model parameters. When
accurate photometric redshifts of strongly lensed systems are
available from the VENUS data (see Section~\ref{sec:venus_data}), we
include the Gaussian priors with a width of typically
$\sigma_z=0.2$. For the cluster MACS~J1149.5+2223, we also include
time delays and flux ratios of multiple images of the supernova Refsdal
\citep[see][]{2025PhRvD.111l3506L,2025PhRvD.112l3526L} and include the
Hubble constant as a free parameter. We compute $\chi^2$ in the source
plane, but taking account of the full magnification tensor as detailed
in Appendix 2 of \citet{2010PASJ...62.1017O}. This method not only
provides an accurate approximation of the image plane $\chi^2$, but
also exhibits excellent performance in a mass modeling challenge
using simulated clusters \citep{2017MNRAS.472.3177M}. Positional
errors of individual multiple images are set to values typically much
larger than measurement errors, considering effects of substructures
in mass distributions, and are adjusted so that the reduced $\chi^2$ of
the best-fitting model becomes roughly equal to one
\citep[see discussions in e.g.,][]{2025OJAp....8E.123N}. Specifically,
typical values of positional errors are $\sim 0\farcs 4-0\farcs 7$,
although we sometimes assume much smaller errors for e.g., knots of
extended arcs or multiple images with small image separations. Errors
on model parameters are derived using the Markov Chain Monte Carlo
(MCMC) methods. There are some exceptional cases (e.g., Abell 209) for
which the reduced $\chi^2$ is much larger than one because of the
small number of observational constraints available for mass modeling,
which forces us to assume a very simple mass model with the small
number of model parameters.

\begin{deluxetable*}{ccrrccccccc} 
  \tabletypesize{\footnotesize}
  \tablewidth{0pt}
  \tablecaption{Summary of {\sc glafic} mass modeling of 60 VENUS clusters. \label{tab:model}}
\tablehead{
  \colhead{Name} & \colhead{$z_{\mathrm{cl}}$} & \colhead{R.A.${}_{\mathrm{cen}}$} & \colhead{Decl.${}_{\mathrm{cen}}$} & \colhead{Analysis region} & \colhead{$N_{\mathrm{img}}$} & \colhead{$N_{\mathrm{sys}}$} & \colhead{$N_{\mathrm{spec}}$} & \colhead{$\chi^2$/dof} & \colhead{$\Delta_{\mathrm{rms}}$} & \colhead{$\sigma_{\mathrm{JK}}$}\\
  & & \colhead{[deg]} & \colhead{[deg]} & \colhead{[${}''$]} &  &  &  &  & \colhead{[${}''$]} & 
}
\startdata
\hline
           Abell~2744 & 0.3080 &   3.5861023 & $-$30.4002210 &  [$-150$, $300$],[$-150$, $300$] & 187 &  63 &  52 & 238.05/212 & 0.47 & 0.99\\
     RXC~J0018.5+1626 & 0.5460 &   4.6398933 &    16.4377372 &  [$-120$, $120$],[$-120$, $120$] &  38 &  15 &   2 &  33.50/24  & 0.58 & 1.29\\
  MACS~J0025.4$-$1222 & 0.5860 &   6.3876685 & $-$12.3879814 &  [$-120$, $120$],[$-120$, $120$] &  19 &   6 &   1 &   6.95/11  & 0.25 & 1.44\\
     RXC~J0032.1+1808 & 0.3956 &   8.0476944 &    18.1409483 &  [$-120$, $120$],[$-180$, $120$] & 160 &  55 &   4 & 190.33/177 & 0.64 & 1.04\\
  MACS~J0035.4$-$2015 & 0.3520 &   8.8589284 & $-$20.2623036 &  [$-120$, $120$],[$-120$, $120$] &  64 &  23 &   7 &  69.34/67  & 0.43 & 0.97\\
             Abell~68 & 0.2550 &   9.2785051 &     9.1567431 &  [$-120$, $120$],[$-120$, $120$] &  42 &  15 &   7 &  30.11/31  & 0.49 & 1.52\\
           Abell~2813 & 0.2920 &  10.8527516 & $-$20.6282542 &  [$-120$, $120$],[$-120$, $130$] &  50 &  16 &   0 &  44.25/41  & 0.51 & 1.04\\
             El~Gordo & 0.8730 &  15.7407763 & $-$49.2720144 &  [$-120$, $120$],[$-120$, $140$] & 164 &  60 &  25 & 153.97/166 & 0.65 & 1.23\\
            Abell~209 & 0.2060 &  22.9689369 & $-$13.6112565 &  [$-120$, $120$],[$-120$, $120$] &   7 &   3 &   1 &   2.96/1   & 0.54 & 0.64\\
   WL~J013719$-$08284 & 0.5660 &  24.3539070 &  $-$8.4563710 &  [$-120$, $120$],[$-120$, $120$] &  26 &   9 &   6 &  25.42/19  & 0.47 & 1.60\\
    CL~J0152.7$-$1357 & 0.8270 &  28.1876184 & $-$13.9510105 &  [$-120$, $120$],[$-120$, $120$] &  30 &  10 &   1 &  15.28/26  & 0.26 & 2.94\\
  MACS~J0159.8$-$0849 & 0.4050 &  29.9555766 &  $-$8.8330430 &  [$-120$, $120$],[$-120$, $120$] &  39 &  17 &   0 &  40.08/33  & 0.58 & 1.08\\
   RXC~J0232.2$-$4420 & 0.2840 &  38.0773351 & $-$44.3466849 &  [$-120$, $120$],[$-120$, $120$] &  45 &  12 &   1 &  50.99/47  & 0.51 & 1.38\\
            Abell~370 & 0.3750 &  39.9696704 &  $-$1.5719347 &  [$-130$, $120$],[$-120$, $120$] & 202 &  68 &  57 & 213.94/219 & 0.66 & 0.83\\
           Abell~S295 & 0.3000 &  41.3534474 & $-$53.0292889 &  [$-130$, $120$],[$-120$, $120$] &  25 &   9 &   4 &  20.49/22  & 0.42 & 1.09\\
            Abell~383 & 0.1870 &  42.0141412 &  $-$3.5291701 &  [$-120$, $120$],[$-120$, $120$] &  30 &   9 &   7 &  35.85/25  & 0.88 & 1.47\\
  MACS~J0257.1$-$2325 & 0.5049 &  44.2865235 & $-$23.4348168 &  [$-120$, $120$],[$-120$, $120$] &  20 &   7 &   1 &  14.72/15  & 0.36 & 1.43\\
  MACS~J0257.6$-$2209 & 0.3220 &  44.4211330 & $-$22.1549437 &  [$-120$, $120$],[$-120$, $120$] &  94 &  27 &  25 & 118.82/117 & 0.68 & 0.97\\
    MACS~J0308.9+2645 & 0.3560 &  47.2332750 &    26.7604961 &  [$-120$, $120$],[$-120$, $120$] &  37 &  16 &   0 &  37.24/31  & 0.46 & 1.19\\
   PLCK~G171.9$-$40.7 & 0.2700 &  48.2370668 &     8.3719286 &  [$-120$, $120$],[$-120$, $120$] &  45 &  14 &   0 &  46.38/46  & 0.50 & 1.29\\
  MACS~J0329.7$-$0211 & 0.4500 &  52.4231973 &  $-$2.1962697 &  [$-120$, $120$],[$-120$, $120$] &  37 &  14 &  10 &  33.00/31  & 0.62 & 1.04\\
           Abell~3192 & 0.4250 &  59.7253342 & $-$29.9253315 &  [$-120$, $120$],[$-120$, $120$] &  52 &  18 &   2 &  52.77/49  & 0.65 & 0.98\\
  MACS~J0416.1$-$2403 & 0.3971 &  64.0381563 & $-$24.0675044 &  [$-120$, $120$],[$-130$, $120$] & 290 & 105 &  94 & 309.01/340 & 0.66 & 0.76\\
  MACS~J0417.5$-$1154 & 0.4430 &  64.3945823 & $-$11.9089217 &  [$-120$, $120$],[$-120$, $120$] & 118 &  40 &  16 & 119.97/134 & 0.58 & 0.91\\
  MACS~J0429.6$-$0253 & 0.3990 &  67.4000869 &  $-$2.8851795 &  [$-120$, $120$],[$-120$, $120$] &  21 &   6 &   3 &  13.39/24  & 0.37 & 0.89\\
    MACS~J0451.9+0006 & 0.4290 &  72.9777295 &     0.1050402 &  [$-120$, $120$],[$-120$, $120$] &  84 &  30 &  25 &  99.91/86  & 0.53 & 0.97\\
            Abell~521 & 0.2480 &  73.5287383 & $-$10.2235399 &  [$-120$, $120$],[$-120$, $120$] &  38 &  16 &  15 &  32.41/35  & 0.16 & 0.62\\
     MS~0451.6$-$0305 & 0.5500 &  73.5451650 &  $-$3.0143277 &  [$-120$, $120$],[$-120$, $120$] &  62 &  22 &   5 &  50.68/52  & 0.44 & 1.48\\
  MACS~J0553.4$-$3342 & 0.4120 &  88.3572793 & $-$33.7077786 &  [$-120$, $180$],[$-120$, $120$] & 147 &  53 &   4 & 165.28/165 & 0.72 & 0.88\\
  MACS~J0600.1$-$2008 & 0.4320 &  90.0340519 & $-$20.1358576 &  [$-280$, $130$],[$-120$, $190$] & 209 &  67 &  20 & 233.51/233 & 0.61 & 1.12\\
       S~J0615$-$5746 & 0.9720 &  93.9656197 & $-$57.7801883 &  [$-120$, $120$],[$-120$, $120$] &  60 &  20 &  12 &  57.30/53  & 0.40 & 1.41\\
       Bullet~cluster & 0.2960 & 104.6587123 & $-$55.9571445 &  [$-120$, $280$],[$-120$, $160$] &  86 &  30 &  24 &  86.39/91  & 0.65 & 0.93\\
    MACS~J0717.5+3745 & 0.5450 & 109.3982478 &    37.7457290 &  [$-120$, $190$],[$-120$, $130$] & 237 &  81 &   7 & 277.05/253 & 0.62 & 1.08\\
     S~J0723.3$-$7327 & 0.3900 & 110.8270090 & $-$73.4547390 &  [$-120$, $120$],[$-120$, $120$] &  50 &  17 &   6 &  33.18/40  & 0.32 & 1.06\\
     RXC~J0911.1+1746 & 0.5049 & 137.7979921 &    17.7747099 &  [$-120$, $120$],[$-120$, $120$] &  11 &   5 &   0 &   1.77/1   & 0.15 & 0.93\\
    MACS~J0940.9+0744 & 0.3350 & 145.2237677 &     7.7403798 &  [$-120$, $120$],[$-120$, $120$] &  25 &   8 &   7 &  26.29/25  & 0.30 & 1.03\\
     RXC~J0949.8+1707$^*$ & 0.3826 & 147.4658585 &    17.1195056 &  [$-120$, $120$],[$-120$, $120$] &  20 &   6 &   2 &   5.24/11  & 0.21 & 0.58\\
    MACS~J1115.8+0129 & 0.3550 & 168.9662572 &     1.4986333 &  [$-120$, $120$],[$-120$, $120$] &  22 &   7 &   1 &  24.01/15  & 0.80 & 1.47\\
     PLCK~G165.7+67.0 & 0.3480 & 171.8196015 &    42.4773058 &  [$-120$, $250$],[$-230$, $150$] & 118 &  41 &  14 &  61.81/121 & 0.07 & 1.18\\
    MACS~J1149.5+2223 & 0.5410 & 177.3987491 &    22.3985308 &  [$-120$, $120$],[$-120$, $120$] & 117 &  38 &  28 & 110.58/126 & 0.45 & 1.01\\
     PLCK~G287.0+32.9 & 0.3830 & 177.7089905 & $-$28.0821928 &  [$-140$, $120$],[$-120$, $120$] &  97 &  31 &  14 &  94.27/99  & 0.68 & 0.97\\
  MACS~J1206.2$-$0847 & 0.4390 & 181.5506031 &  $-$8.8009300 &  [$-120$, $120$],[$-120$, $120$] & 124 &  41 &  37 & 128.32/139 & 0.52 & 0.78\\
  MACS~J1311.0$-$0310 & 0.4940 & 197.7575102 &  $-$3.1777062 &  [$-120$, $120$],[$-120$, $120$] &  13 &   4 &   1 &  14.33/8   & 0.73 & 2.16\\
           Abell~1689 & 0.1870 & 197.8729233 &  $-$1.3410994 &  [$-140$, $120$],[$-120$, $130$] & 165 &  55 &  17 & 155.08/183 & 0.55 & 0.99\\
   RXC~J1347.5$-$1145$^*$ & 0.4510 & 206.8775419 & $-$11.7526347 &  [$-120$, $120$],[$-120$, $120$] & 129 &  40 &  38 & 138.17/157 & 0.65 & 0.89\\
           Abell~1835$^*$ & 0.2530 & 210.2586212 &     2.8784684 &  [$-120$, $120$],[$-120$, $120$] &  21 &   7 &   2 &  21.97/8   & 0.79 & 3.89\\
    MACS~J1423.8+2404 & 0.5450 & 215.9494781 &    24.0784501 &  [$-120$, $120$],[$-120$, $120$] &  27 &   8 &   5 &  36.23/27  & 0.64 & 1.42\\
     RXC~J1532.9+3021 & 0.3630 & 233.2240780 &    30.3498152 &  [$-120$, $120$],[$-120$, $120$] &  16 &   6 &   0 &   8.71/13  & 0.28 & 0.76\\
   PLCK~G004.5$-$19.5 & 0.5160 & 289.2711616 & $-$33.5223393 &  [$-120$, $120$],[$-120$, $120$] &  50 &  17 &   1 &  55.36/48  & 0.51 & 2.23\\
  MACS~J1931.8$-$2635 & 0.3520 & 292.9567621 & $-$26.5757431 &  [$-120$, $120$],[$-120$, $120$] &  40 &  15 &   8 &  39.54/37  & 0.49 & 1.03\\
     S~J2031.8$-$4036 & 0.3420 & 307.9719397 & $-$40.6252203 &  [$-120$, $120$],[$-120$, $120$] &  66 &  22 &  19 &  78.47/73  & 0.43 & 0.97\\
  MACS~J2129.4$-$0741 & 0.5890 & 322.3587881 &  $-$7.6910536 &  [$-120$, $120$],[$-120$, $120$] &  63 &  20 &  12 &  74.99/67  & 0.63 & 1.10\\
     RXC~J2129.7+0005 & 0.2340 & 322.4164769 &     0.0892336 &  [$-120$, $120$],[$-120$, $120$] &  38 &  14 &  11 &  21.99/37  & 0.37 & 0.86\\
     S~J2131.1$-$4019 & 0.4420 & 322.7705195 & $-$40.3225398 &  [$-120$, $120$],[$-120$, $120$] &  27 &  10 &   9 &  13.36/9   & 0.47 & 1.64\\
       MS~2137$-$2353 & 0.3130 & 325.0632041 & $-$23.6611906 &  [$-120$, $120$],[$-120$, $120$] &  22 &   8 &   4 &  13.37/20  & 0.31 & 0.81\\
   RXC~J2211.7$-$0350 & 0.3970 & 332.9413169 &  $-$3.8290195 &  [$-120$, $120$],[$-120$, $120$] & 146 &  43 &   3 & 174.99/182 & 0.61 & 1.10\\
  MACS~J2214.9$-$1359 & 0.5020 & 333.7386868 & $-$14.0035677 &  [$-120$, $120$],[$-120$, $120$] &  56 &  18 &  16 &  54.54/57  & 0.65 & 1.02\\
          Abell~S1063 & 0.3480 & 342.1831870 & $-$44.5308979 &  [$-120$, $120$],[$-120$, $120$] & 173 &  62 &  27 & 205.60/193 & 0.42 & 1.20\\
           Abell~2537 & 0.2966 & 347.0926033 &  $-$2.1921637 &  [$-120$, $120$],[$-120$, $120$] &  57 &  15 &   1 &  74.55/67  & 0.69 & 1.06\\
SPT$-$CL~J2325$-$4111 & 0.3580 & 351.2988390 & $-$41.2035767 &  [$-120$, $120$],[$-120$, $120$] &  56 &  16 &   6 &  60.27/57  & 0.59 & 1.13
\enddata
\tablecomments{Parameters listed in the table include the cluster
  redshift $z_{\mathrm{cl}}$, the reference coordinates
  (R.A.${}_{\mathrm{cen}}$, Decl.${}_{\mathrm{cen}}$), the size of the
  region [$x_{\mathrm{min}}$, $x_{\mathrm{max}}$],
  [$y_{\mathrm{min}}$, $y_{\mathrm{max}}$] ($x$ and $y$ are relative
  coordinates from reference coordinates with the West being $+x$ and
  North being $+y$) for the analysis of 
  the lensing strength in Section~\ref{sec:strength}, the number
  of multiple images $N_{\mathrm{img}}$, the number of strongly lensed
  systems $N_{\mathrm{sys}}$, the number of strongly lensed systems
  with spectroscopic redshifts $N_{\mathrm{spec}}$, the $\chi^2$ of
  the best-fitting model and the number of degrees of freedom (dof),
  the root-mean-square of differences between observed and
  model-predicted multiple image positions $\Delta_{\mathrm{rms}}$,
  and the standard deviation of the Jackknife distribution
  $\sigma_{\mathrm{JK}}$ with $4\sigma$ clipping. Asterisks in the
  cluster names indicate that JWST VENUS images are not yet available
  and therefore are not used for mass modeling of these clusters.} 
\end{deluxetable*}

\begin{center}
    \begin{figure*}[!t]
        \centering
    	\includegraphics[width=0.75\linewidth]{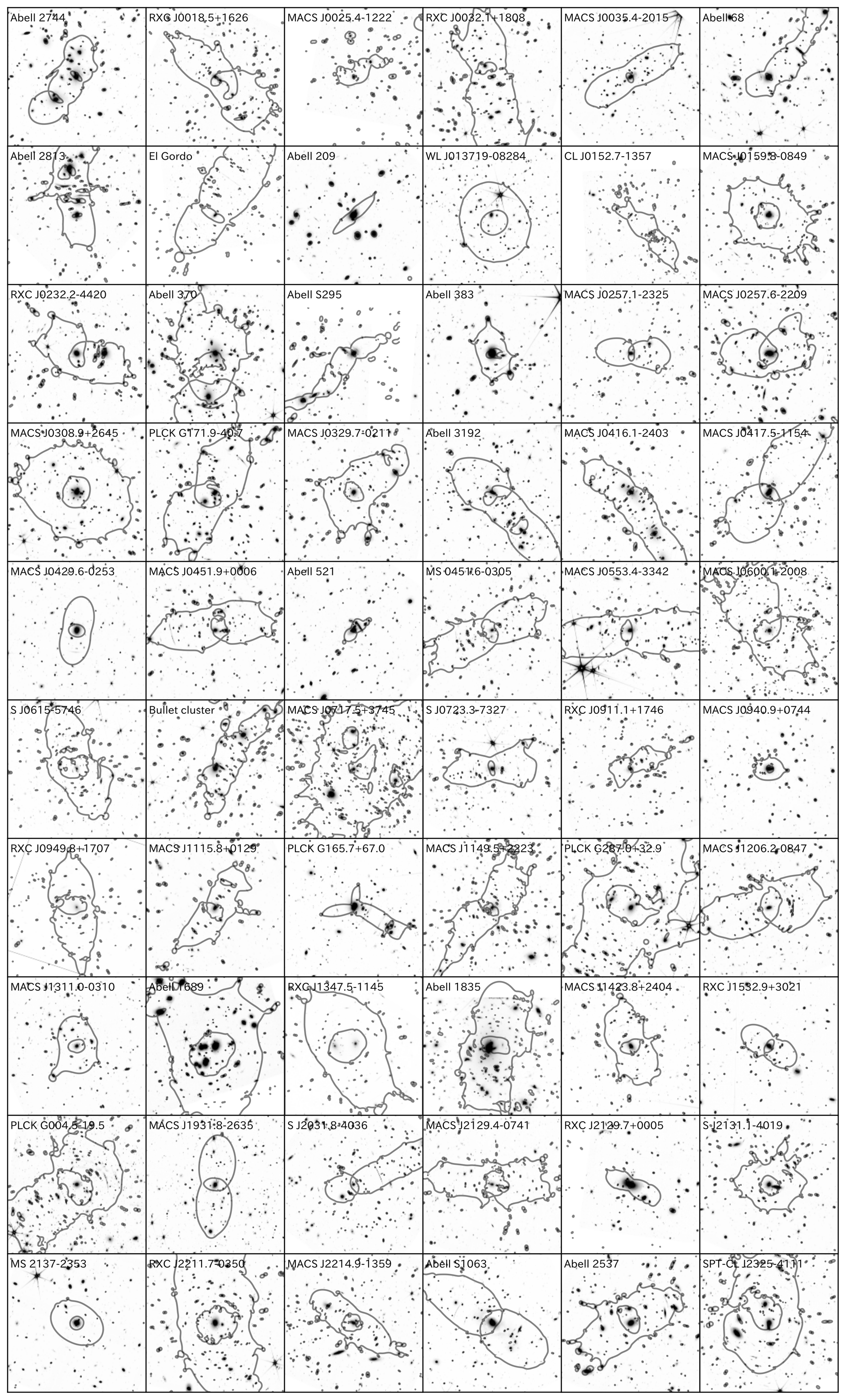}
    	\caption{JWST F200W images of 60 VENUS clusters
          with critical curves of best-fitting models for the source
          redshift $z_{\mathrm{s}}=9$ overlaid. North is up and east
          is to the left. Each panel has the size of $120''\times
          120''$ and is centered at the reference coordinates listed in
        Table~\ref{tab:model}.}
        \label{fig:jwst_images}
    \end{figure*}
\end{center}

\section{Mass Modeling Results}
\label{sec:model}

Table~\ref{tab:model} summarizes the information on the input multiple
images as well as $\chi^2$ and the number of the degrees of freedom for
the best-fitting model for all the 60 clusters.\footnote{We will make
lens maps publicly available as part of DR1 together
with other VENUS products such as JWST images and object catalogs.}
Table~\ref{tab:model} also presents the root-mean-square (rms) of
differences between observed multiple image positions
$\boldsymbol{\theta}_i^{\mathrm{obs}}$ and model-predicted multiple
image positions $\boldsymbol{\theta}_i^{\mathrm{model}}$ defined by
\begin{equation}
  \Delta_{\mathrm{rms}}=\sqrt{\frac{1}{N_{\mathrm{img}}}\sum_{i=1}^{N_{\mathrm{img}}}\left|\boldsymbol{\theta}_i^{\mathrm{obs}}-\boldsymbol{\theta}_i^{\mathrm{model}}\right|^2},
\end{equation}
where $N_{\mathrm{img}}$ is the number of multiple images used for mass
modeling of each cluster. We show JWST images and critical curves of
best-fitting models in Figure~\ref{fig:jwst_images}.

\begin{center}
    \begin{figure}[!t]
        \centering
    	\includegraphics[scale=0.42]{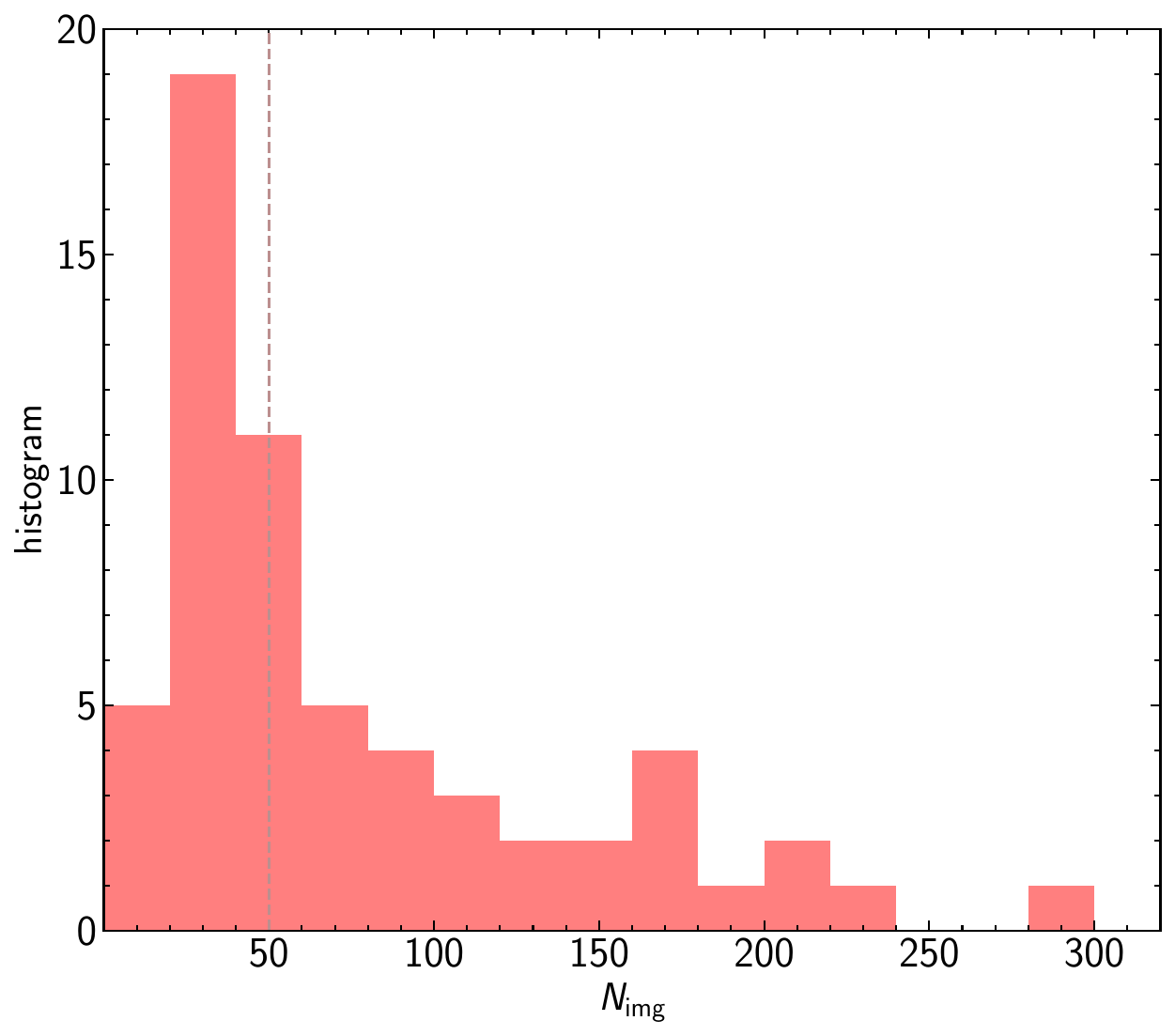}
    	\caption{The histogram of the numbers of multiple images
          $N_{\mathrm{img}}$. The vertical dashed line indicates the
          median value.}
        \label{fig:hist_nimg}
    \end{figure}
\end{center}

Figure~\ref{fig:hist_nimg} shows the histogram of
$N_{\mathrm{img}}$. It is found that each cluster typically contains
$N_{\mathrm{img}}\sim 50$ multiple images, and about $25\%$ of
clusters contain more than 100 multiple images. The large number
of multiple images is crucial for tight constraints on mass
distributions of individual galaxy clusters.

\begin{center}
    \begin{figure}[!t]
        \centering
    	\includegraphics[scale=0.42]{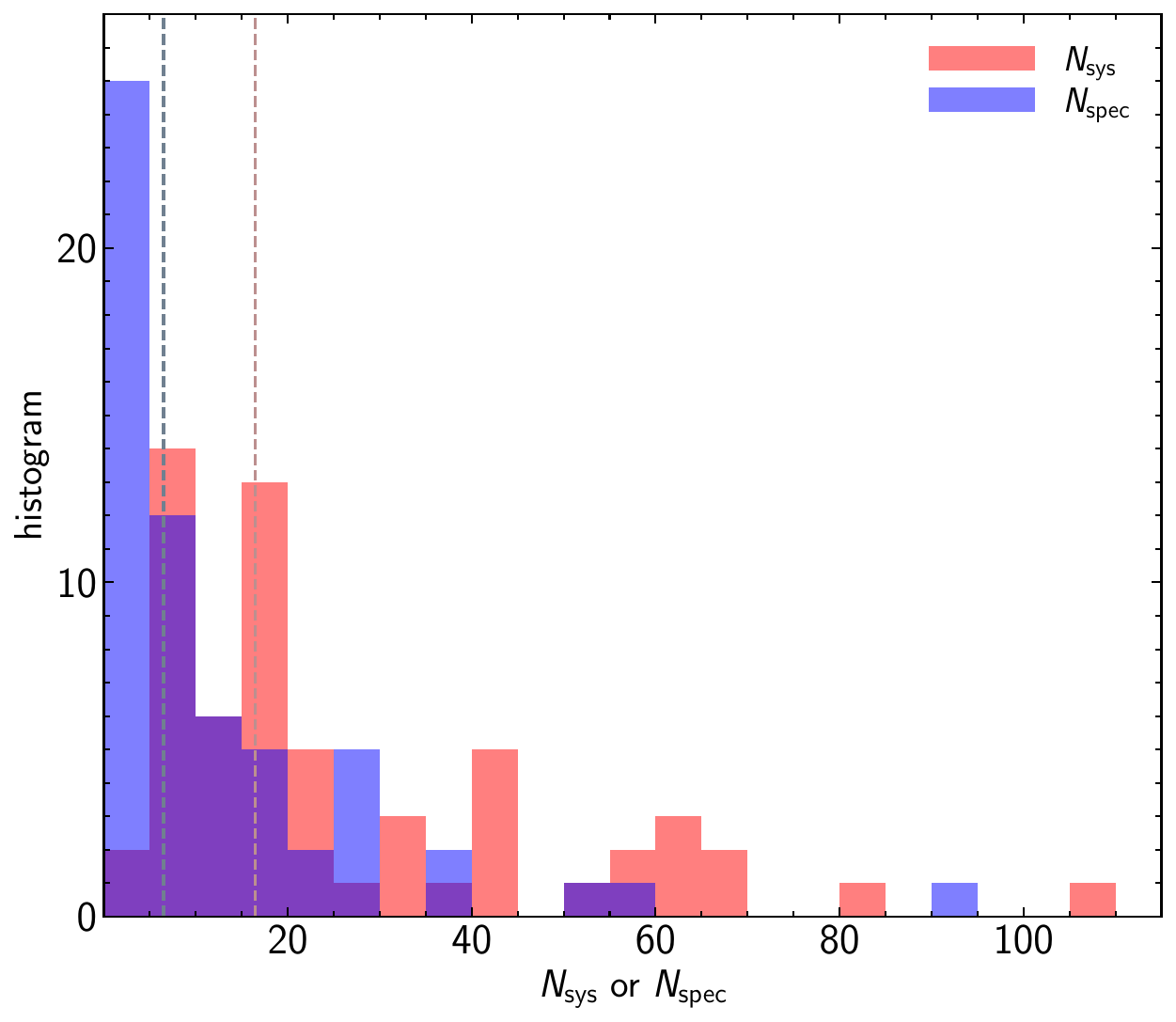}
    	\caption{The histograms of the numbers of strongly lensed
          systems $N_{\mathrm{sys}}$ and those with spectroscopic
          redshifts $N_{\mathrm{spec}}$. The vertical dashed lines
          indicate the median values.} 
        \label{fig:hist_nsys}
    \end{figure}
\end{center}

Figure~\ref{fig:hist_nsys} shows the histograms of the numbers of
strongly lensed systems $N_{\mathrm{sys}}$ and those with
spectroscopic redshifts $N_{\mathrm{spec}}$. Since $\sim 2-4$ multiple
images are typically produced for each strongly lensed background
galaxy, $N_{\mathrm{sys}}$ is roughly one-third of $N_{\mathrm{img}}$.
On the other hand, $N_{\mathrm{spec}}$ varies significantly across
clusters (see also Table~\ref{tab:model}), reflecting the fact that
spectroscopic redshifts are obtained by various follow-up
observations, typically with the Multi Unit Spectroscopic Explorer
(MUSE) on the Very Large Telescope
\citep[e.g.,][]{2019A&A...632A..36C,2021A&A...646A..83R}, the HST
\citep[e.g.,][]{2015ApJ...812..114T}, and the JWST
\citep[e.g.,][]{2023MNRAS.525.1867N,2025ApJ...987..186F,2026ApJS..282....3S}. 
We note that, as part of VENUS, spectroscopy with the JWST
micro-shutter assembly (MSA) for 10 out of the 60 clusters is planned,
which will increase significantly $N_{\mathrm{spec}}$ for these clusters.

\begin{center}
    \begin{figure}[!t]
        \centering
    	\includegraphics[scale=0.42]{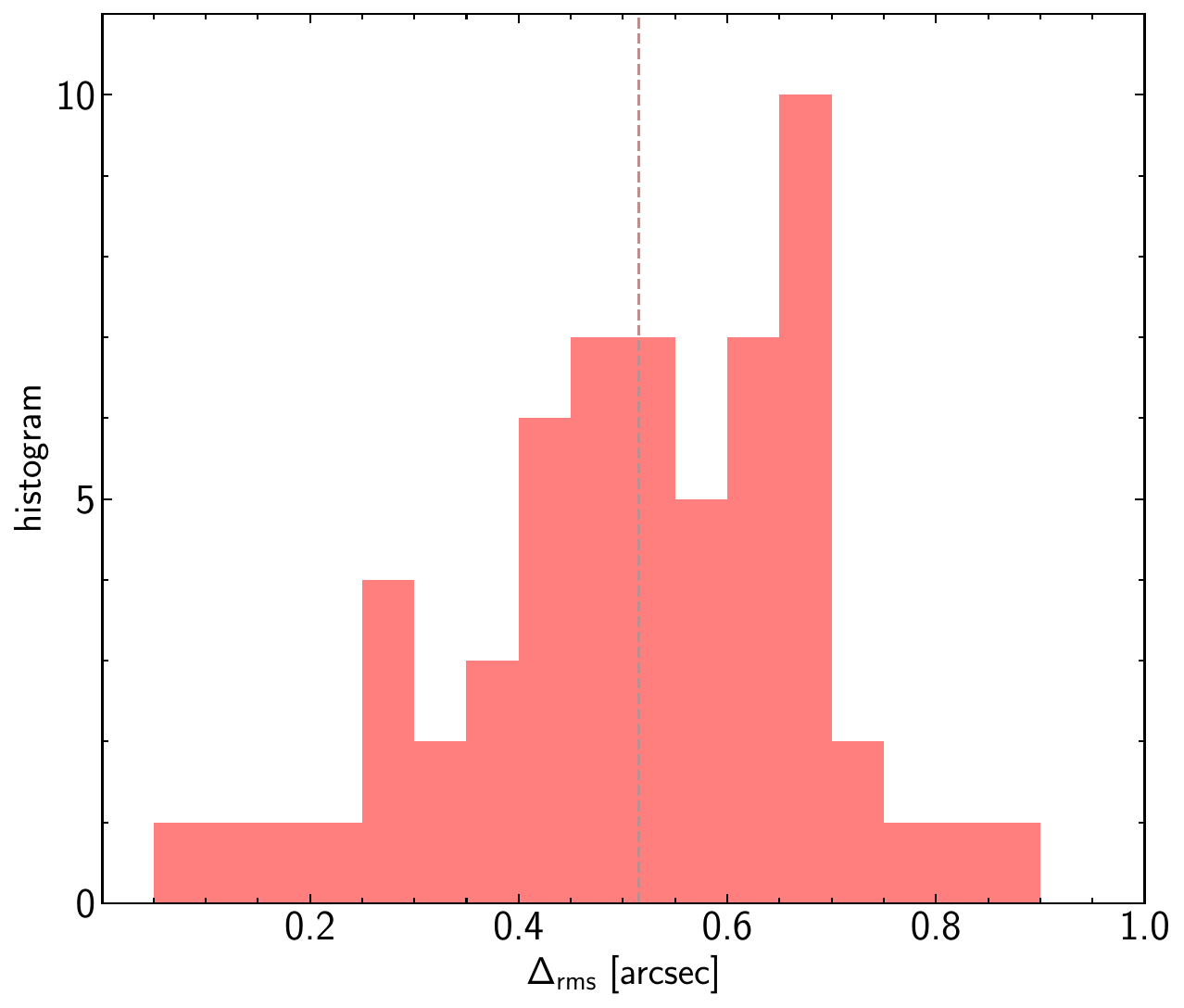}
    	\caption{The histograms of the root-mean-square of differences
          between observed and model-predicted multiple image
          positions $\Delta_{\mathrm{rms}}$. The vertical dashed line
          indicates the median value.} 
        \label{fig:hist_rms}
    \end{figure}
\end{center}

Figure~\ref{fig:hist_rms} shows the histogram of
$\Delta_{\mathrm{rms}}$, which is often used to quantify the precision
of a mass model. We find that $\Delta_{\mathrm{rms}}$ of our mass
models is typically $\sim 0\farcs5$, which is comparable to the rms of
other parametric lens models using the similar number of multiple
images \citep[see e.g.,][for a recent compilation of the rms for HFF mass
  models]{2023ApJ...951..140C}. All our mass models have $\Delta_{\mathrm{rms}}$
smaller than $1''$.

\begin{center}
    \begin{figure}[!t]
        \centering
    	\includegraphics[scale=0.42]{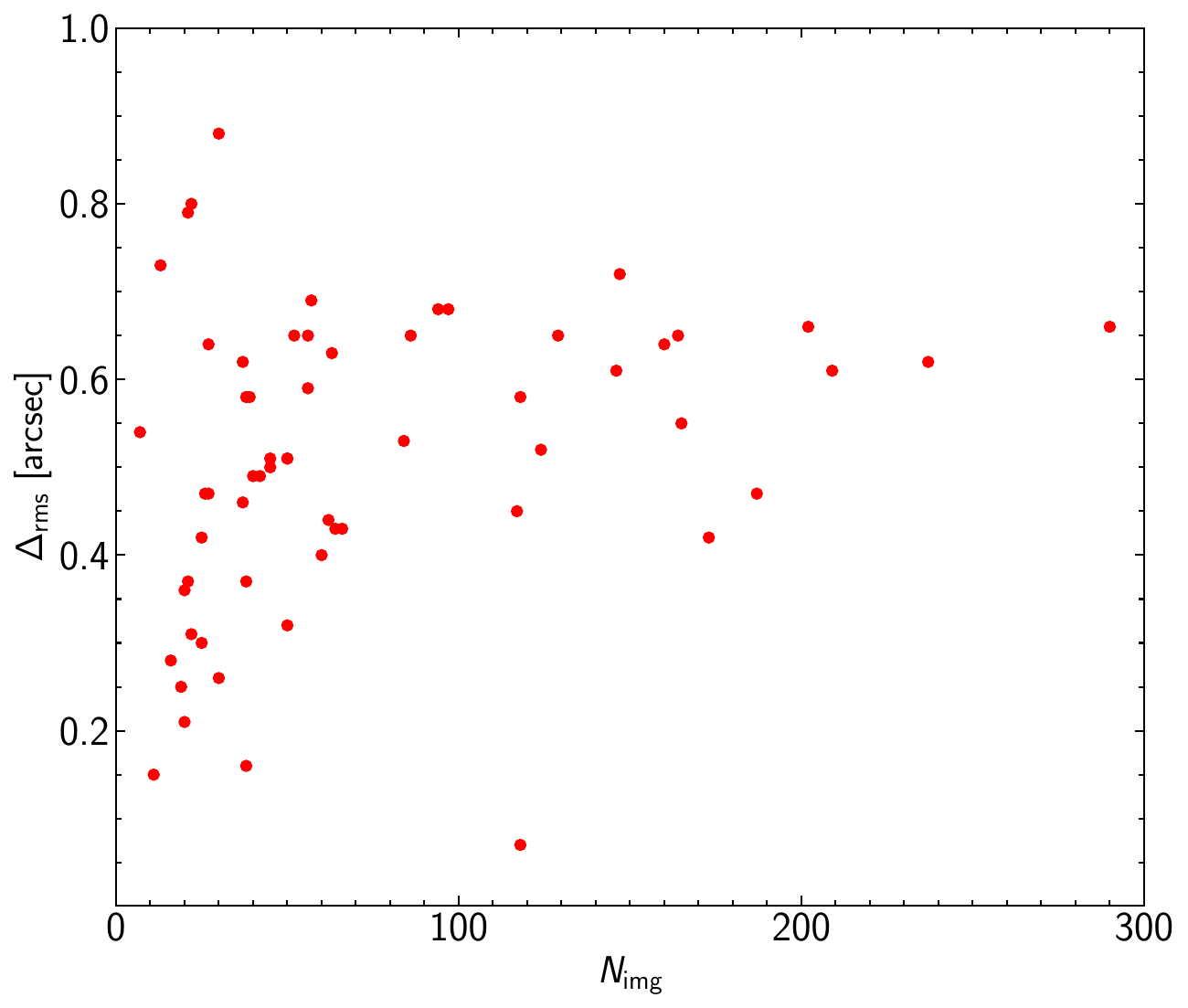}
    	\caption{The distribution of the numbers of multiple images
          $N_{\mathrm{img}}$ and the root-mean-square of differences
          between observed and model-predicted multiple image
          positions $\Delta_{\mathrm{rms}}$.}  
        \label{fig:nimg_rms}
    \end{figure}
\end{center}

While the value of $\Delta_{\mathrm{rms}}$ mainly reflects the
complexity of the cluster mass distribution that is not fully captured
by simple parametric mass modeling, it may also be affected by
overfitting of multiple image positions
\citep[see e.g.,][]{2025OJAp....8E.123N}. In
Figure~\ref{fig:nimg_rms}, we check the possible dependence of
$\Delta_{\mathrm{rms}}$ on $N_{\mathrm{img}}$. We find that
$\Delta_{\mathrm{rms}}$ does not strongly depend on
$N_{\mathrm{img}}$, although there appears a weak trend of increasing
$\Delta_{\mathrm{rms}}$ with increasing $N_{\mathrm{img}}$. This trend
may partly be due to overfitting of multiple image positions when
$N_{\mathrm{img}}$ is small. An alternative explanation is larger
$\Delta_{\mathrm{rms}}$ for more massive clusters with larger Einstein
radii. Figure~\ref{fig:nimg_rms} also indicates that PLCK~G165.7+67.0
is an outlier in the $\Delta_{\mathrm{rms}}$--$N_{\mathrm{img}}$
relation with the very small rms $\Delta_{\mathrm{rms}}\simeq 0.07$ for $N_{\mathrm{img}}=118$,
which is because many knots in lensed giant arcs are included as
constraints in order to constrain $H_0$ with the lensed supernova H0pe
as precisely as possible \citep{2025ApJ...979...13P}.

\begin{center}
    \begin{figure}[!t]
        \centering
    	\includegraphics[scale=0.42]{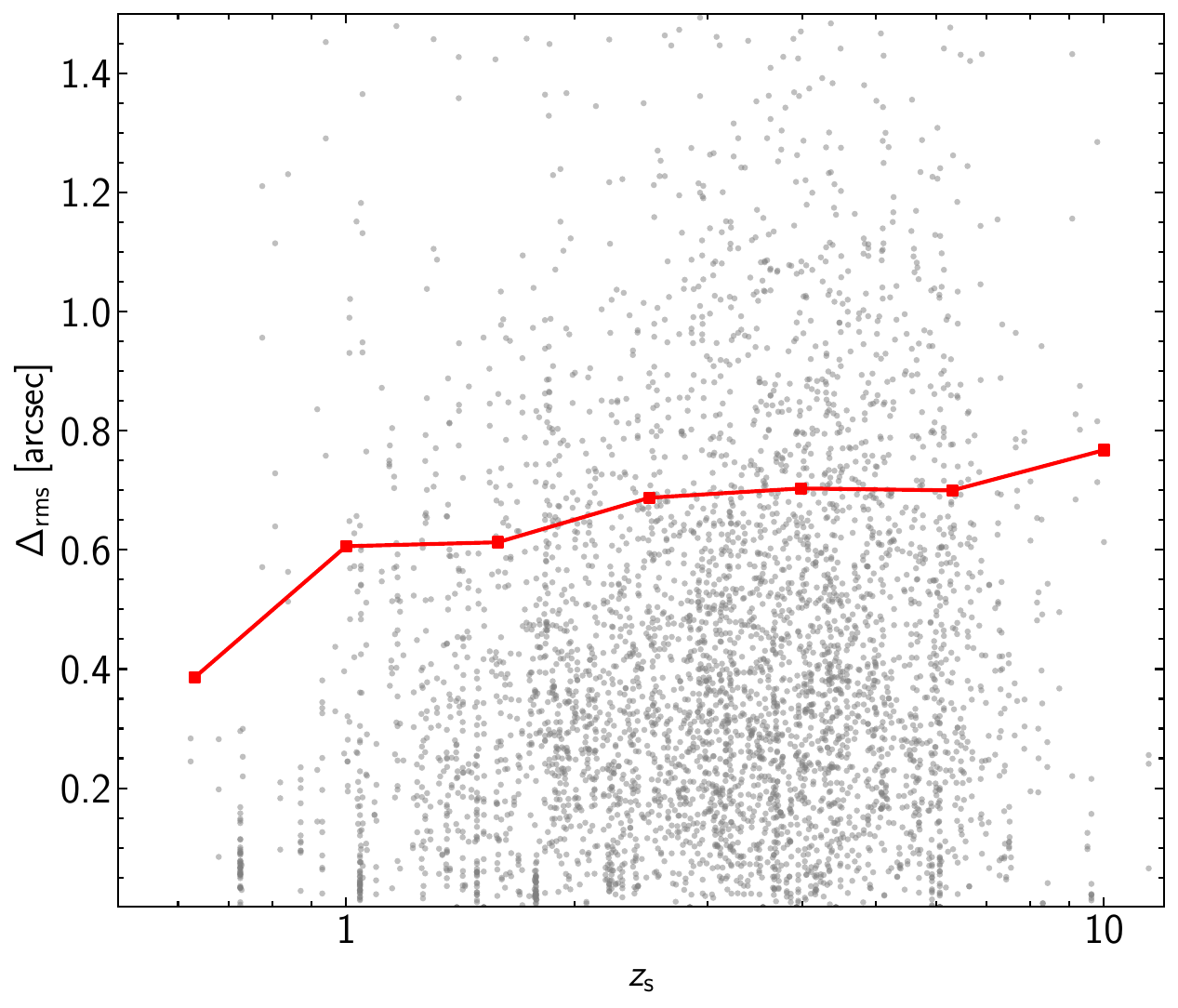}
    	\caption{The root-mean-square of differences
          between observed and model-predicted multiple image
          $\Delta_{\mathrm{rms}}$ as a function of the source redshift
        $z_{\mathrm{s}}$, where $z_{\mathrm{s}}$ are taken from values
          obtained by the best-fitting mass model. We use the bin
          width of $\Delta(\log z_{\mathrm{s}})=0.2$ to derive 
          $\Delta_{\mathrm{rms}}$ for each bin. The offsets for
          individual multiple images are shown by gray circles.}  
        \label{fig:zs_rms}
    \end{figure}
\end{center}

We also check the dependence of $\Delta_{\mathrm{rms}}$ on the source
redshift $z_{\mathrm{s}}$, which may become important for studying
physical properties of strongly lensed high-redshift galaxies. The
result shown in Figure~\ref{fig:zs_rms} suggests that there is a
trend of increasing $\Delta_{\mathrm{rms}}$ with increasing
$z_{\mathrm{s}}$, although the trend is weak. More quantitatively,
$\Delta_{\mathrm{rms}}\simeq 0\farcs6$ for $z_{\mathrm{s}}\simeq 1-2$,
and increases to $\simeq 0\farcs 7$ for $z_{\mathrm{s}}\gtrsim 3$ and
to  $\simeq 0\farcs 8$ for $z_{\mathrm{s}}\gtrsim 10$. This trend is
presumably because source galaxies at higher redshifts are typically
located in the outer parts of multiple image regions where constraints on
the mass models are less stringent due to small numbers of multiple
images in those regions. 

\begin{center}
    \begin{figure}[!t]
        \centering
    	\includegraphics[scale=0.42]{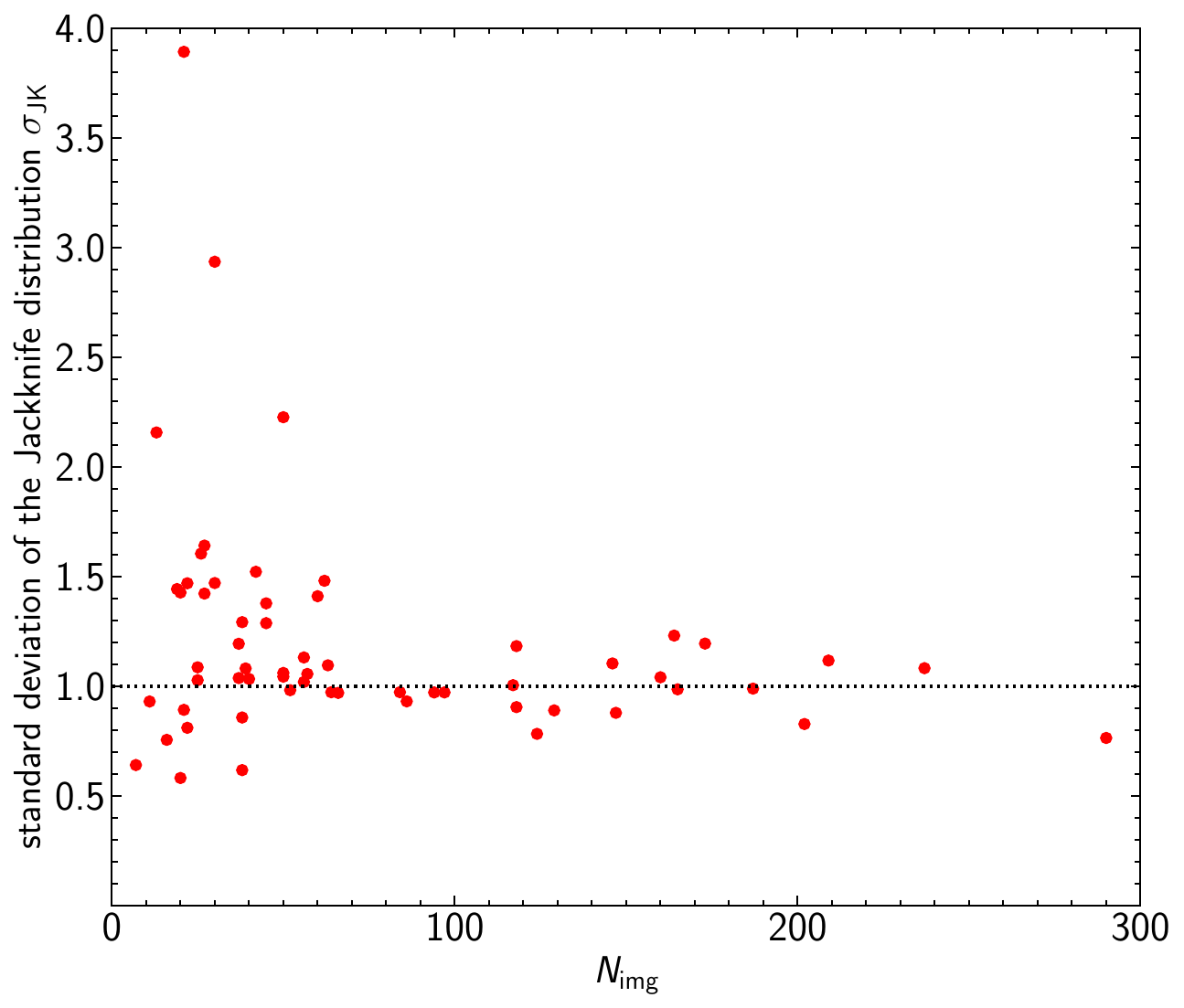}
    	\caption{The distribution of the numbers of multiple images
          $N_{\mathrm{img}}$ and the standard deviation of the
          Jackknife distribution $\sigma_{\mathrm{JK}}$. The
          horizontal dotted line shows $\sigma_{\mathrm{JK}}=1$, which
        is an expected value of $\sigma_{\mathrm{JK}}$ when the mass
        model does not suffer from overfitting.} 
        \label{fig:nimg_jk}
    \end{figure}
\end{center}

\citet{2025OJAp....8E.123N} propose the Jackknife method, for which
best-fitting models with one of multiple image systems removed are
used to check the predictive power of mass models, as a method of
checking the accuracy of mass models
\citep[see also][for its application to the time-delay cosmography]{2025PhRvD.112l3526L}.
In the Jackknife method, the degree of ``correctness'' of a mass model
is quantified by the standard deviation of the so-called Jackknife
distribution, which refers to a histogram of ratios of differences
between observed and model-predicted positions of removed images to
assumed positional errors of those images \citep[see][for more
  details]{2025OJAp....8E.123N}.  
The derived standard deviation $\sigma_{\mathrm{JK}}$ with $4\sigma$
clipping to mitigate effects of e.g., isolated multiple image systems
for each cluster is listed in
Table~\ref{tab:model}. Figure~\ref{fig:nimg_jk} shows the dependence
of $\sigma_{\mathrm{JK}}$ on $N_{\mathrm{img}}$. We find 
that $\sigma_{\mathrm{JK}}\sim 1$ for most of the clusters, indicating
no sign of overfitting. This is particularly true for clusters with
large $N_{\mathrm{img}}$, which is consistent with an argument in
\citet{2025OJAp....8E.123N} that a large number of multiple images is
crucial for mitigating the overfitting. We find that there are a few
clusters with $\sigma_{\mathrm{JK}}> 2$ for which our current mass
models are likely overfitting multiple image positions. In fact, the
cluster with the highest $\sigma_{\mathrm{JK}}$ is Abell 1835 for
which the JWST NIRCam imaging is lacking. We expect that new JWST
NIRCam images for this cluster will enable us to identify more
multiple images that help reduce $\sigma_{\mathrm{JK}}$.

\begin{deluxetable*}{cccccccccc} 
  \tabletypesize{\footnotesize}
  \tablewidth{0pt}
  \tablecaption{High magnification areas $\mathcal{A}_{|\mu|>3}$, $\mathcal{A}_{|\mu|>10}$, and the Einstein radii $\theta_{\mathrm{Ein}}$ of 60 VENUS clusters.\label{tab:area_tein}}
\tablehead{
    & \multicolumn{3}{c}{$\mathcal{A}_{|\mu|>3}$ [arcmin$^2$]} & & $\mathcal{A}_{|\mu|>10}$ [arcmin$^2$] & &\multicolumn{3}{c}{$\theta_{\mathrm{Ein}}$ [arcsec]} \\
\cline{2-4} \cline{6-6} \cline{8-10}
\colhead{Name} & \colhead{$z_{\mathrm{s}}=1$} & \colhead{$z_{\mathrm{s}}=3$} & \colhead{$z_{\mathrm{s}}=9$} & & \colhead{$z_{\mathrm{s}}=9$} & &\colhead{$z_{\mathrm{s}}=1$} & \colhead{$z_{\mathrm{s}}=3$} & \colhead{$z_{\mathrm{s}}=9$}
}
\startdata
\hline
           Abell~2744 & $1.921\pm0.077$ & $4.206\pm0.211$ & $4.989\pm0.271$ & & $1.071\pm0.043$ & & $16.56\pm 0.10$ & $23.58\pm 0.08$ & $26.53\pm 0.14$ \\
     RXC~J0018.5+1626 & $0.634\pm0.042$ & $3.020\pm0.335$ & $4.006\pm0.461$ & & $1.220\pm0.191$ & & $ 7.67\pm 0.89$ & $24.85\pm 0.40$ & $31.85\pm 0.35$ \\
  MACS~J0025.4$-$1222 & $0.231\pm0.022$ & $1.734\pm0.403$ & $2.491\pm0.539$ & & $0.521\pm0.124$ & & $ 3.71\pm 0.74$ & $11.91\pm 0.38$ & $15.08\pm 0.45$ \\
     RXC~J0032.1+1808 & $2.809\pm0.097$ & $5.149\pm0.245$ & $5.972\pm0.301$ & & $1.771\pm0.112$ & & $23.15\pm 0.57$ & $43.46\pm 0.26$ & $48.31\pm 0.22$ \\
  MACS~J0035.4$-$2015 & $0.862\pm0.029$ & $1.861\pm0.091$ & $2.269\pm0.127$ & & $0.565\pm0.025$ & & $12.93\pm 0.24$ & $21.77\pm 0.18$ & $24.93\pm 0.21$ \\
             Abell~68 & $2.315\pm0.242$ & $3.958\pm0.508$ & $4.621\pm0.609$ & & $1.200\pm0.119$ & & $ 8.11\pm 0.40$ & $22.18\pm 0.89$ & $26.32\pm 0.34$ \\
           Abell~2813 & $2.116\pm0.247$ & $4.016\pm0.658$ & $4.772\pm0.824$ & & $1.119\pm0.132$ & & $19.20\pm 0.94$ & $27.34\pm 0.69$ & $30.62\pm 0.61$ \\
             El~Gordo & $0.006\pm0.001$ & $3.216\pm0.116$ & $4.645\pm0.229$ & & $1.366\pm0.086$ & & $ 0.65\pm 0.07$ & $30.78\pm 0.56$ & $44.80\pm 0.53$ \\
            Abell~209 & $0.370\pm0.134$ & $0.565\pm0.241$ & $0.639\pm0.286$ & & $0.139\pm0.043$ & & $ 7.21\pm 0.65$ & $ 9.42\pm 0.69$ & $10.32\pm 0.78$ \\
   WL~J013719$-$08284 & $0.453\pm0.030$ & $1.520\pm0.403$ & $1.945\pm0.597$ & & $0.531\pm0.194$ & & $11.01\pm 2.05$ & $26.52\pm 0.34$ & $31.56\pm 0.49$ \\
    CL~J0152.7$-$1357 & $0.006\pm0.002$ & $1.026\pm0.134$ & $1.757\pm0.219$ & & $0.426\pm0.059$ & & $ 0.35\pm 0.05$ & $13.00\pm 0.31$ & $19.07\pm 0.44$ \\
  MACS~J0159.8$-$0849 & $1.194\pm0.087$ & $2.629\pm0.280$ & $3.226\pm0.356$ & & $0.961\pm0.141$ & & $14.86\pm 0.66$ & $27.64\pm 0.21$ & $32.40\pm 0.46$ \\
   RXC~J0232.2$-$4420 & $1.618\pm0.118$ & $2.704\pm0.284$ & $3.133\pm0.344$ & & $0.915\pm0.098$ & & $19.03\pm 0.48$ & $27.66\pm 0.14$ & $30.19\pm 0.22$ \\
            Abell~370 & $2.330\pm0.032$ & $4.325\pm0.096$ & $5.052\pm0.131$ & & $1.822\pm0.037$ & & $21.52\pm 0.10$ & $37.77\pm 0.13$ & $43.94\pm 0.21$ \\
           Abell~S295 & $2.071\pm0.344$ & $4.110\pm0.836$ & $4.853\pm1.022$ & & $1.103\pm0.181$ & & $ 7.67\pm 0.32$ & $17.38\pm 1.46$ & $21.10\pm 1.19$ \\
            Abell~383 & $1.011\pm0.169$ & $1.412\pm0.268$ & $1.558\pm0.306$ & & $0.466\pm0.077$ & & $13.42\pm 0.31$ & $17.43\pm 0.21$ & $18.84\pm 0.24$ \\
  MACS~J0257.1$-$2325 & $0.315\pm0.016$ & $1.310\pm0.129$ & $1.757\pm0.192$ & & $0.434\pm0.042$ & & $ 6.17\pm 0.25$ & $17.01\pm 0.25$ & $20.79\pm 0.30$ \\
  MACS~J0257.6$-$2209 & $1.436\pm0.050$ & $2.697\pm0.132$ & $3.198\pm0.161$ & & $0.906\pm0.043$ & & $17.19\pm 0.16$ & $27.51\pm 0.14$ & $30.60\pm 0.18$ \\
    MACS~J0308.9+2645 & $2.016\pm0.092$ & $3.555\pm0.245$ & $4.055\pm0.305$ & & $1.080\pm0.100$ & & $26.48\pm 0.38$ & $40.09\pm 0.14$ & $44.11\pm 0.28$ \\
   PLCK~G171.9$-$40.7 & $2.606\pm0.247$ & $4.374\pm0.584$ & $5.075\pm0.709$ & & $1.284\pm0.139$ & & $23.51\pm 0.46$ & $34.11\pm 0.22$ & $37.44\pm 0.26$ \\
  MACS~J0329.7$-$0211 & $0.959\pm0.038$ & $2.332\pm0.097$ & $2.883\pm0.127$ & & $0.850\pm0.046$ & & $11.31\pm 0.32$ & $26.78\pm 0.32$ & $31.22\pm 0.42$ \\
           Abell~3192 & $1.332\pm0.083$ & $2.931\pm0.270$ & $3.563\pm0.364$ & & $1.127\pm0.106$ & & $18.63\pm 0.43$ & $30.81\pm 0.45$ & $35.16\pm 0.55$ \\
  MACS~J0416.1$-$2403 & $1.063\pm0.015$ & $1.967\pm0.040$ & $2.271\pm0.050$ & & $0.629\pm0.016$ & & $15.98\pm 0.15$ & $28.36\pm 0.17$ & $32.30\pm 0.14$ \\
  MACS~J0417.5$-$1154 & $1.411\pm0.025$ & $3.824\pm0.148$ & $4.701\pm0.200$ & & $1.147\pm0.037$ & & $14.07\pm 0.11$ & $34.09\pm 0.16$ & $39.51\pm 0.16$ \\
  MACS~J0429.6$-$0253 & $0.393\pm0.030$ & $0.794\pm0.100$ & $0.942\pm0.127$ & & $0.271\pm0.037$ & & $10.91\pm 0.45$ & $17.98\pm 0.16$ & $20.32\pm 0.22$ \\
    MACS~J0451.9+0006 & $0.927\pm0.018$ & $2.044\pm0.069$ & $2.467\pm0.093$ & & $0.673\pm0.022$ & & $11.59\pm 0.33$ & $28.13\pm 0.17$ & $32.01\pm 0.16$ \\
            Abell~521 & $0.382\pm0.144$ & $0.944\pm0.462$ & $1.210\pm0.587$ & & $0.160\pm0.042$ & & $ 5.20\pm 0.14$ & $ 8.25\pm 0.25$ & $ 9.53\pm 0.55$ \\
     MS~0451.6$-$0305 & $0.663\pm0.036$ & $3.697\pm0.522$ & $5.116\pm0.700$ & & $1.391\pm0.192$ & & $ 6.55\pm 0.73$ & $24.92\pm 0.20$ & $31.87\pm 0.37$ \\
  MACS~J0553.4$-$3342 & $2.038\pm0.086$ & $4.509\pm0.278$ & $5.531\pm0.354$ & & $1.452\pm0.093$ & & $14.99\pm 0.55$ & $37.00\pm 0.14$ & $42.64\pm 0.15$ \\
  MACS~J0600.1$-$2008 & $4.165\pm0.065$ & $8.429\pm0.192$ & $9.781\pm0.228$ & & $3.308\pm0.088$ & & $13.03\pm 0.28$ & $53.31\pm 0.38$ & $62.62\pm 0.22$ \\
       S~J0615$-$5746 & $0.000\pm0.000$ & $1.385\pm0.046$ & $2.280\pm0.113$ & & $0.678\pm0.029$ & & $ 0.23\pm 0.00$ & $23.49\pm 0.12$ & $32.96\pm 0.19$ \\
       Bullet~cluster & $2.913\pm0.260$ & $6.922\pm0.755$ & $8.363\pm0.945$ & & $1.411\pm0.132$ & & $12.70\pm 0.69$ & $24.05\pm 0.29$ & $26.75\pm 0.24$ \\
    MACS~J0717.5+3745 & $2.725\pm0.025$ & $6.328\pm0.108$ & $7.306\pm0.169$ & & $2.373\pm0.067$ & & $ 8.90\pm 0.63$ & $57.22\pm 0.19$ & $65.37\pm 0.17$ \\
     S~J0723.3$-$7327 & $1.039\pm0.097$ & $2.780\pm0.431$ & $3.610\pm0.569$ & & $0.909\pm0.093$ & & $ 9.05\pm 0.18$ & $20.59\pm 0.25$ & $25.36\pm 0.38$ \\
     RXC~J0911.1+1746 & $0.171\pm0.018$ & $0.744\pm0.149$ & $1.033\pm0.245$ & & $0.250\pm0.054$ & & $ 5.58\pm 0.96$ & $13.77\pm 0.84$ & $17.28\pm 0.68$ \\
    MACS~J0940.9+0744 & $0.147\pm0.034$ & $0.244\pm0.086$ & $0.281\pm0.107$ & & $0.071\pm0.028$ & & $ 7.49\pm 0.59$ & $10.69\pm 0.08$ & $11.53\pm 0.12$ \\
     RXC~J0949.8+1707$^*$ & $0.940\pm0.088$ & $1.879\pm0.321$ & $2.224\pm0.427$ & & $0.630\pm0.118$ & & $15.54\pm 1.22$ & $25.80\pm 0.57$ & $29.72\pm 0.18$ \\
    MACS~J1115.8+0129 & $0.929\pm0.103$ & $1.864\pm0.366$ & $2.265\pm0.459$ & & $0.629\pm0.113$ & & $14.21\pm 0.75$ & $22.23\pm 0.51$ & $25.24\pm 0.58$ \\
     PLCK~G165.7+67.0 & $0.519\pm0.026$ & $1.214\pm0.112$ & $1.538\pm0.160$ & & $0.383\pm0.026$ & & $ 4.78\pm 0.33$ & $14.84\pm 0.63$ & $17.49\pm 0.75$ \\
    MACS~J1149.5+2223 & $0.747\pm0.178$ & $3.542\pm0.729$ & $4.647\pm0.939$ & & $1.410\pm0.354$ & & $ 6.81\pm 0.24$ & $24.55\pm 1.19$ & $34.98\pm 4.16$ \\
     PLCK~G287.0+32.9 & $4.042\pm0.090$ & $7.937\pm0.335$ & $9.337\pm0.365$ & & $3.211\pm0.126$ & & $22.32\pm 0.52$ & $49.26\pm 0.34$ & $57.75\pm 0.28$ \\
  MACS~J1206.2$-$0847 & $1.184\pm0.014$ & $2.851\pm0.051$ & $3.515\pm0.076$ & & $1.038\pm0.019$ & & $15.12\pm 0.12$ & $30.24\pm 0.10$ & $35.65\pm 0.14$ \\
  MACS~J1311.0$-$0310 & $0.323\pm0.029$ & $0.971\pm0.129$ & $1.225\pm0.176$ & & $0.370\pm0.069$ & & $ 7.72\pm 0.66$ & $17.58\pm 0.41$ & $21.52\pm 0.49$ \\
           Abell~1689 & $4.725\pm0.175$ & $5.922\pm0.274$ & $6.336\pm0.301$ & & $1.815\pm0.081$ & & $43.22\pm 0.19$ & $51.19\pm 0.17$ & $53.91\pm 0.20$ \\
   RXC~J1347.5$-$1145$^*$ & $1.350\pm0.022$ & $2.956\pm0.104$ & $3.498\pm0.138$ & & $0.961\pm0.051$ & & $20.59\pm 0.37$ & $36.90\pm 0.09$ & $42.03\pm 0.12$ \\
           Abell~1835$^*$ & $2.680\pm0.394$ & $4.163\pm0.703$ & $4.714\pm0.835$ & & $1.345\pm0.224$ & & $26.19\pm 0.63$ & $35.91\pm 0.49$ & $39.38\pm 0.61$ \\
    MACS~J1423.8+2404 & $0.376\pm0.015$ & $1.526\pm0.111$ & $2.017\pm0.171$ & & $0.599\pm0.044$ & & $ 7.17\pm 0.25$ & $21.35\pm 0.21$ & $26.68\pm 0.25$ \\
     RXC~J1532.9+3021 & $0.556\pm0.068$ & $1.378\pm0.269$ & $1.735\pm0.375$ & & $0.429\pm0.070$ & & $ 7.82\pm 0.65$ & $15.55\pm 0.22$ & $18.36\pm 0.20$ \\
   PLCK~G004.5$-$19.5 & $1.414\pm0.033$ & $3.657\pm0.118$ & $4.392\pm0.161$ & & $1.429\pm0.067$ & & $17.29\pm 0.45$ & $39.29\pm 0.21$ & $46.37\pm 0.20$ \\
  MACS~J1931.8$-$2635 & $0.951\pm0.026$ & $1.857\pm0.073$ & $2.208\pm0.096$ & & $0.598\pm0.024$ & & $14.74\pm 0.22$ & $23.92\pm 0.13$ & $26.89\pm 0.14$ \\
     S~J2031.8$-$4036 & $1.347\pm0.213$ & $2.312\pm0.381$ & $2.650\pm0.459$ & & $0.741\pm0.137$ & & $14.31\pm 1.94$ & $29.28\pm 2.24$ & $32.31\pm 2.53$ \\
  MACS~J2129.4$-$0741 & $0.420\pm0.016$ & $2.281\pm0.161$ & $3.170\pm0.282$ & & $0.916\pm0.057$ & & $ 7.56\pm 0.22$ & $24.38\pm 0.22$ & $31.44\pm 0.40$ \\
     RXC~J2129.7+0005 & $0.697\pm0.074$ & $1.191\pm0.154$ & $1.395\pm0.189$ & & $0.297\pm0.027$ & & $ 9.77\pm 0.14$ & $13.76\pm 0.17$ & $15.21\pm 0.24$ \\
     S~J2131.1$-$4019 & $0.632\pm0.020$ & $1.514\pm0.091$ & $1.816\pm0.121$ & & $0.575\pm0.057$ & & $10.34\pm 0.52$ & $22.83\pm 0.31$ & $26.57\pm 0.22$ \\
       MS~2137$-$2353 & $0.438\pm0.031$ & $0.670\pm0.061$ & $0.745\pm0.072$ & & $0.191\pm0.018$ & & $14.09\pm 0.12$ & $18.97\pm 0.15$ & $20.54\pm 0.21$ \\
   RXC~J2211.7$-$0350 & $2.409\pm0.046$ & $4.277\pm0.123$ & $4.884\pm0.142$ & & $1.453\pm0.043$ & & $30.08\pm 0.49$ & $46.47\pm 0.25$ & $51.97\pm 0.26$ \\
  MACS~J2214.9$-$1359 & $0.532\pm0.028$ & $1.643\pm0.134$ & $2.155\pm0.192$ & & $0.604\pm0.049$ & & $ 9.58\pm 0.25$ & $21.58\pm 0.29$ & $25.86\pm 0.36$ \\
          Abell~S1063 & $1.975\pm0.039$ & $3.770\pm0.084$ & $4.430\pm0.108$ & & $1.269\pm0.030$ & & $20.00\pm 0.10$ & $33.29\pm 0.38$ & $38.43\pm 0.32$ \\
           Abell~2537 & $1.854\pm0.122$ & $3.061\pm0.236$ & $3.474\pm0.287$ & & $1.041\pm0.096$ & & $19.68\pm 0.57$ & $29.47\pm 0.28$ & $32.97\pm 0.24$ \\
SPT$-$CL~J2325$-$4111 & $2.762\pm0.176$ & $6.166\pm0.538$ & $7.566\pm0.709$ & & $2.130\pm0.158$ & & $19.55\pm 0.63$ & $36.45\pm 0.25$ & $42.42\pm 0.41$ 
 \enddata
\tablecomments{Asterisks in the cluster names indicate
  that JWST VENUS images are not yet available and therefore are
   not used for mass modeling of these clusters.}
\end{deluxetable*}

\section{Quantifying Lensing Strengths}
\label{sec:strength}

\subsection{Area of the High Magnification Region}

The area of the high magnification region in the image plane is often
adopted as an indicator of the lensing strength of each cluster
\citep[e.g.,][]{2014MNRAS.444..268R,2022ApJ...928...87F}. In this
paper, we mainly employ $\mathcal{A}_{|\mu|>3}$, which is defined by the
image-plane area in which a source at redshift $z_{\mathrm{s}}$ is
magnified by a factor of 3 or more, to quantify the lensing strength.
The threshold magnification value of 3 is chosen to ensure sufficient
magnifications for studying high-redshift galaxies. 
The analysis is conducted within the region shown in
Table~\ref{tab:model}. For all the clusters, we assume the minimum
analysis region size of $240''\times 240''$ centered at the brightest
cluster galaxy to make sure that it fully covers the high
magnification region associated with the main cluster halo. For some
of the clusters, the analysis region is extended further to fully
include JWST images of interest. For each cluster, we estimate the
mean and the error of $\mathcal{A}_{|\mu|>3}$ from the mean and scatter of
$\mathcal{A}_{|\mu|>3}$ values for 50 mass models that are randomly drawn
from the MCMC chain. In addition to $\mathcal{A}_{|\mu|>3}$, for
the source redshift of $z_{\mathrm{s}}=9$, we also check the area for
the higher magnification threshold, $\mathcal{A}_{|\mu|>10}$.
Table~\ref{tab:area_tein} summarizes $\mathcal{A}_{|\mu|>3}$
for the source redshift $z_{\mathrm{s}}=1$, $3$, and $9$, as
well as $\mathcal{A}_{|\mu|>10}$ for $z_{\mathrm{s}}=9$.
For $z_{\mathrm{s}}=9$, we identify PLCK~G287.0+32.9,
  MACS~J0600.1$-$2008, and Bullet~cluster as clusters with the
largest $\mathcal{A}_{|\mu|>3}$ among our sample.
 
\subsection{Einstein Radius}

Another quantity that is commonly used to quantify the lensing
strength of each cluster is the Einstein radius
\citep[e.g.,][]{2011MNRAS.410.1939Z,2022ApJ...928...87F}. The
definition of the Einstein radius is, however, not unique when the
mass distribution is not circularly symmetric on the sky. Throughout the
paper, we employ the Einstein radius $\theta_{\mathrm{Ein}}$ defined
by the area enclosed by the tangential critical curve. Specifically,
within the analysis region shown in Table~\ref{tab:model}, we search
for the largest contiguous region defined by $1-\kappa-\gamma<0$,
where $\kappa$ and $\gamma$ are convergence and shear. Given the area
$\mathcal{A}_{1-\kappa-\gamma<0}$ of the largest contiguous region,
the Einstein radius is defined by
\begin{equation}
  \theta_{\mathrm{Ein}}=\sqrt{\frac{\mathcal{A}_{1-\kappa-\gamma<0}}{\pi}}.
  \label{eq:def_tein}
\end{equation}
In the same manner as for $\mathcal{A}_{|\mu|>3}$, we derive
$\theta_{\mathrm{Ein}}$ and its error from 50 mass models that are
randomly drawn from the MCMC chain. Table~\ref{tab:area_tein}
summarizes $\theta_{\mathrm{Ein}}$ values for the source redshift
$z_{\mathrm{s}}=1$, $3$, and $9$. For $z_{\mathrm{s}}=9$, we identify
MACS~J0717.5+3745, MACS~J0600.1$-$2008, and PLCK~G287.0+32.9 as
clusters with the largest $\theta_{\mathrm{Ein}}$ among the VENUS
clusters
\citep[see also][]{2009ApJ...707L.102Z,2017ApJ...839L..11Z,2024MNRAS.533.2242F}. 

\begin{center}
    \begin{figure}[!t]
        \centering
    	\includegraphics[scale=0.42]{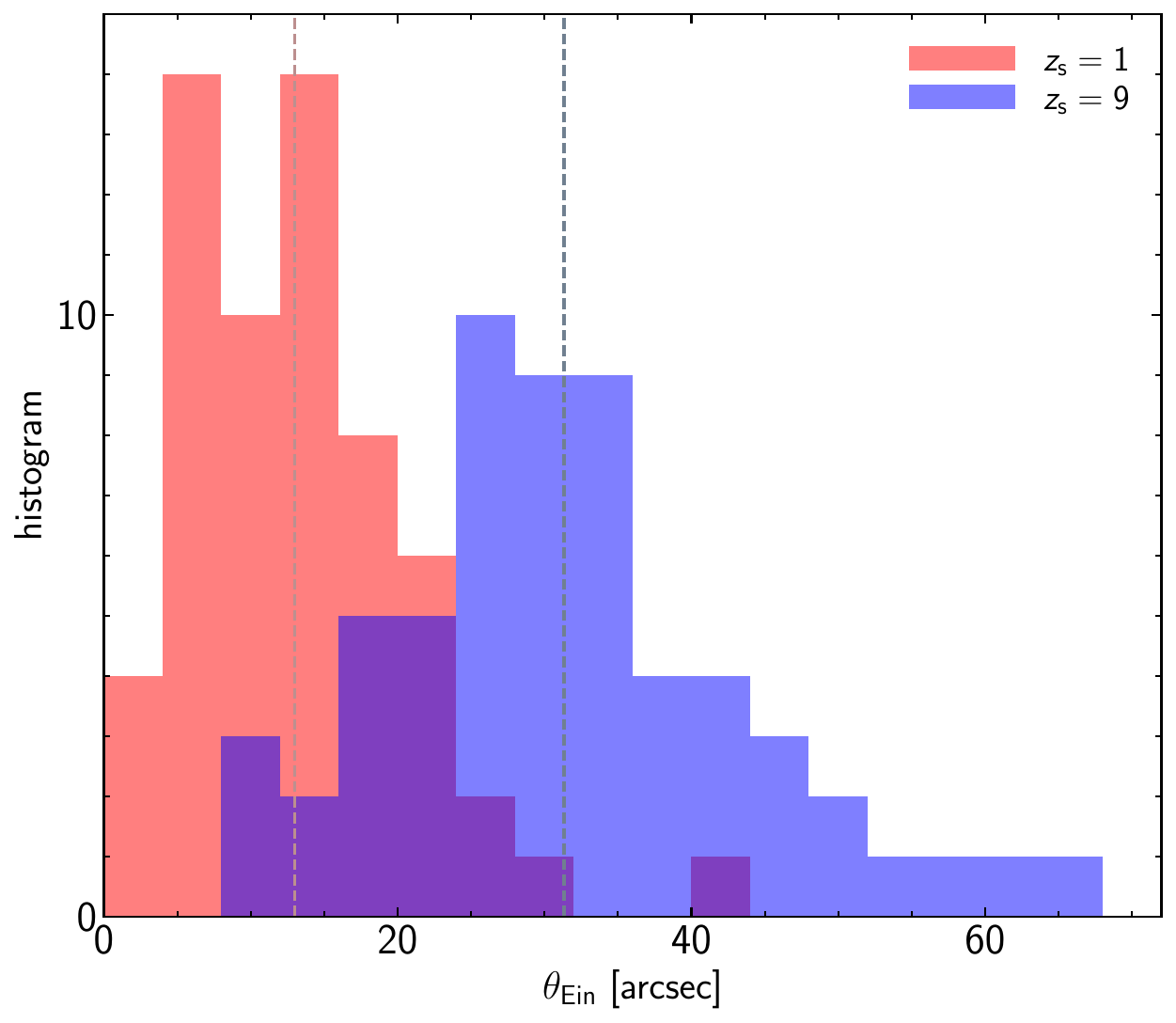}
    	\caption{The histogram of the Einstein radii for the source
          redshifts of $z_{\mathrm{s}}=1$ ({\it red}) and $9$ ({\it
            blue}). The vertical dashed lines
          indicate the median values.}
        \label{fig:hist_tein}
    \end{figure}
\end{center}

We show the histograms of the Einstein radii for the source redshifts
of $z_{\mathrm{s}}=1$ and $9$ in Figure~\ref{fig:hist_tein}. The
Einstein radii are larger for the higher source redshift, because of
the increase of the deflection angle as $\propto
D_{\mathrm{ls}}/D_{\mathrm{os}}$, where $D_{\mathrm{ls}}$ and
$D_{\mathrm{os}}$ are angular diameter distances from the lens and
the source and from the observer to the source, respectively.
We find that the VENUS clusters typically have the Einstein radii
$\theta_{\mathrm{Ein}}\sim 30''$ for $z_{\mathrm{s}}=9$, which confirms
that they have large Einstein radii on average. This is not surprising
because VENUS selects target clusters based on their high lensing
strengths. 

\subsection{Correlation of the High-Magnification Areas and the Einstein Radii}

\begin{center}
\begin{figure*}[!t]
        \centering
    	\includegraphics[scale=0.42]{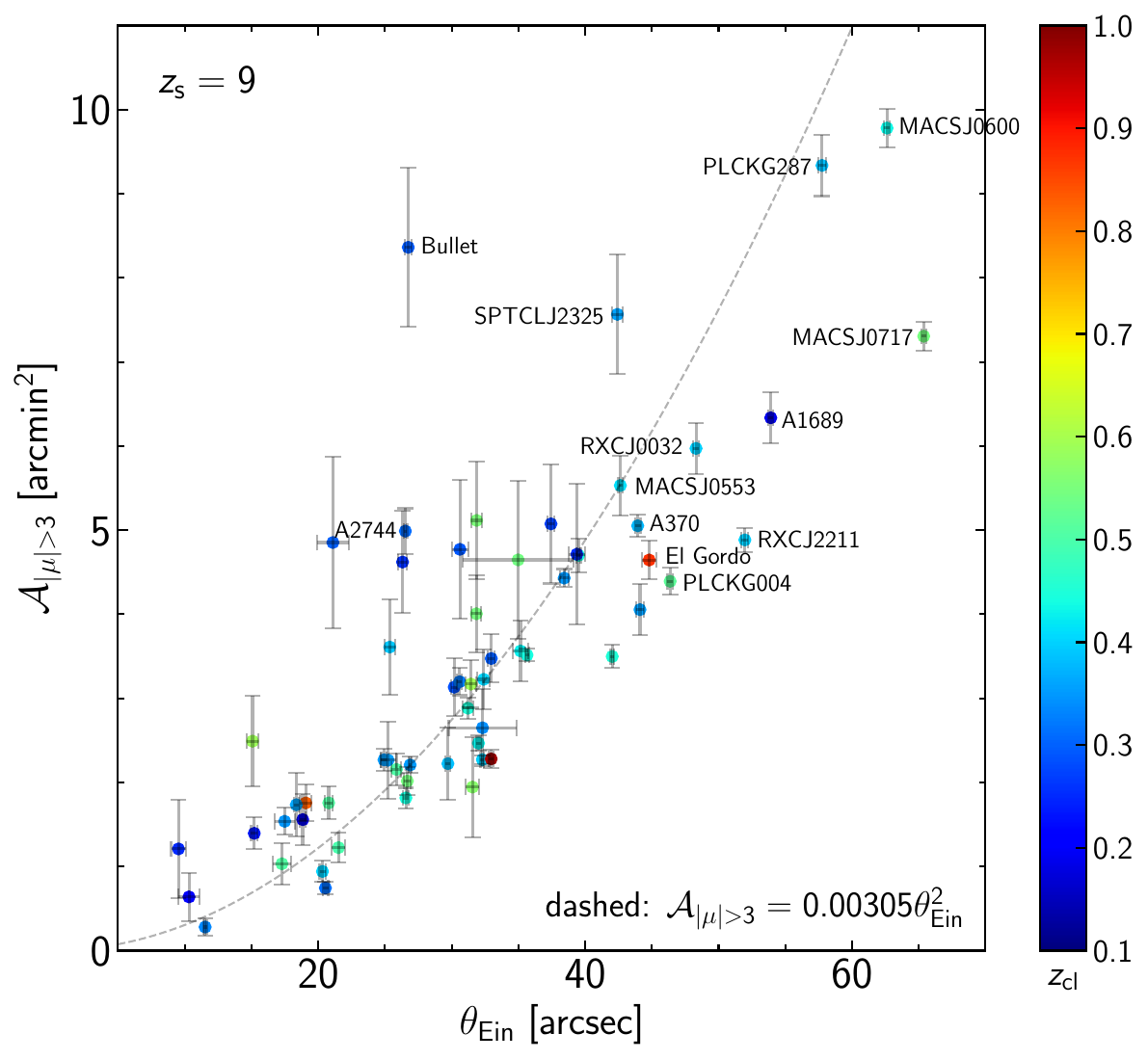}
        \includegraphics[scale=0.42]{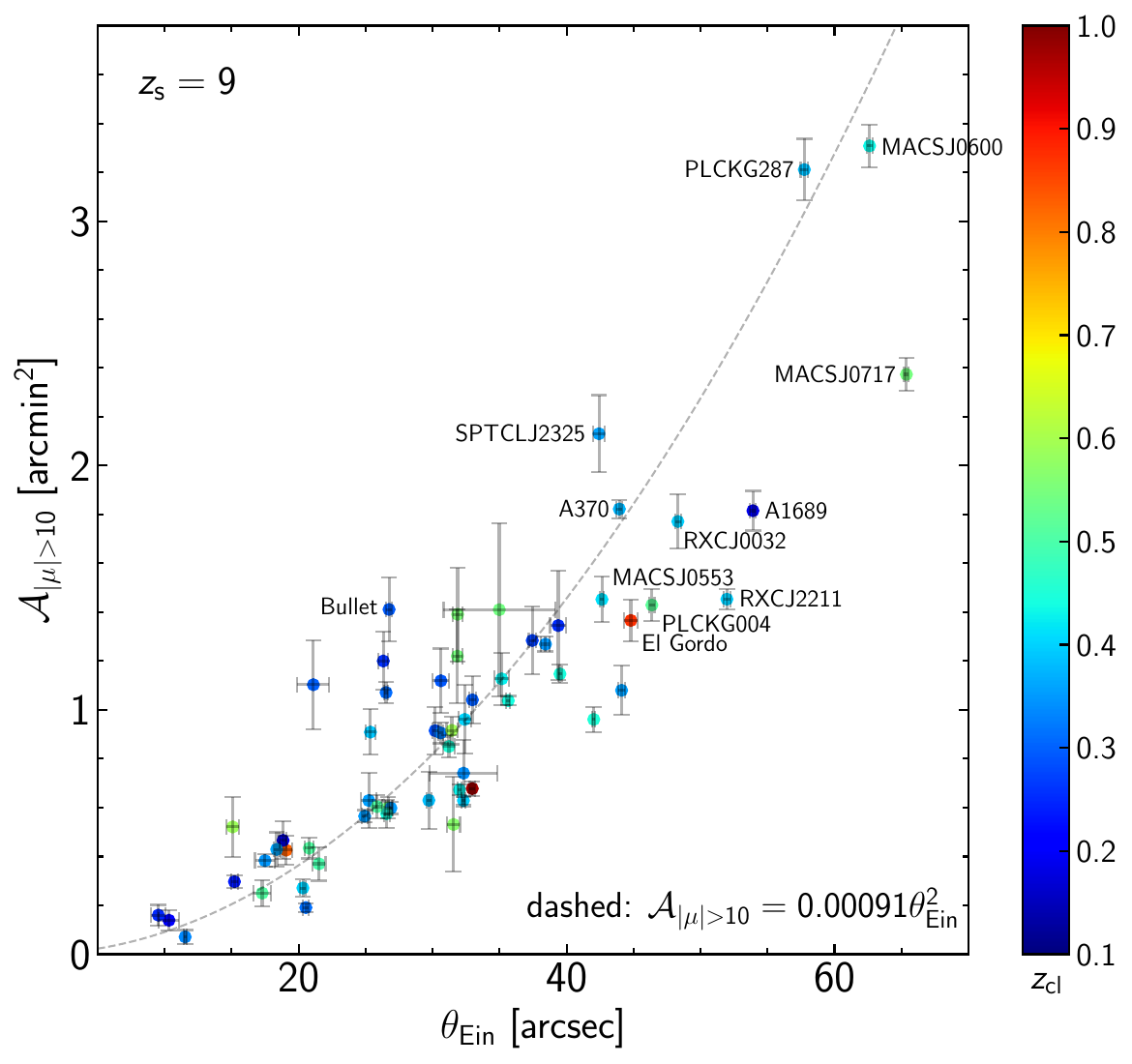}
    	\caption{{\it Left:} The distribution of $\mathcal{A}_{|\mu|>3}$ and
          $\theta_{\mathrm{Ein}}$, both of which quantify lensing
          strengths, for the source redshift of
          $z_{\mathrm{s}}=9$. The color variation is based on the
          cluster redshift $z_{\mathrm{cl}}$. Error bars show
          $1\sigma$ errors. Their values are also summarized in
          Table~\ref{tab:area_tein}. The dashed line is a median fit
          assuming $\mathcal{A}_{|\mu|>3}\propto
          \theta_{\mathrm{Ein}}^2$. {\it Right:} Similar to the left
          panel, but for $\mathcal{A}_{|\mu|>10}$. 
          }  
        \label{fig:area_tein}
    \end{figure*}
\end{center}

Since both $\mathcal{A}_{|\mu|>3}$ and $\theta_{\mathrm{Ein}}$
quantify lensing strengths, we naturally expect a correlation of
these two quantities. To explicitly check for such a correlation, in
the left panel of Figure~\ref{fig:area_tein} we show the distribution of
$\mathcal{A}_{|\mu|>3}$ and $\theta_{\mathrm{Ein}}$ for the source
redshift of $z_{\mathrm{s}}=9$. We find that $\mathcal{A}_{|\mu|>3}$
and $\theta_{\mathrm{Ein}}$ are indeed positively correlated with each
other with roughly $\mathcal{A}_{|\mu|>3}\propto \theta_{\mathrm{Ein}}^2$.
Some clusters such as Bullet~cluster and Abell~2744 are
outliers of the correlation in the sense that they have relatively
large $\mathcal{A}_{|\mu|>3}$ compared with values expected from their
$\theta_{\mathrm{Ein}}$. This is simply because of the existence of
large sub-clumps in these clusters; our definitions of
$\mathcal{A}_{|\mu|>3}$ and $\theta_{\mathrm{Ein}}$ are in such a way
that these sub-clumps significantly increase $\mathcal{A}_{|\mu|>3}$
while they do not significantly affect $\theta_{\mathrm{Ein}}$ that is
solely determined by the size of the critical curve of the main clump
in our definition. In addition, the large external shear from other clumps
can enhance the area of the high-magnification region in each clump
for such complex systems. We caution that some clusters required two
pointings to cover the critical curves (see the analysis region column in
Table~\ref{tab:model}), which indicates that $\mathcal{A}_{|\mu|>3}$
for some of the clusters cannot be well covered by just one JWST
NIRCam pointing.  

In the right panel of Figure~\ref{fig:area_tein}, we also show a
plot similar to the left panel of Figure~\ref{fig:area_tein}, but for
the area of the high-magnification region with the higher
magnification threshold, $\mathcal{A}_{|\mu|>10}$. This Figure
suggests that the area is better correlated with the Einstein radius
$\theta_{\mathrm{Ein}}$ such that
$\mathcal{A}_{|\mu|>10}\propto \theta_{\mathrm{Ein}}^2$ for this 
higher magnification threshold of $|\mu|>10$. This Figure also
confirms our previous argument that, among the 60 VENUS clusters,
MACS~J0717.5+3745, MACS~J0600.1$-$2008, and PLCK~G287.0+32.9, which
also have the largest Einstein radii for $z_{\mathrm{s}}=9$, are the
best natural telescopes to study high-redshift Universe in terms of
their large high-magnification areas. On the other hand, clusters
with large $\mathcal{A}_{|\mu|>3}$, such as Bullet, probe a
complementary parameter space and therefore are also important for
studying high-redshift Universe .

\section{The Largest Einstein Radius in the Universe}
\label{sec:ein}

The distribution of Einstein radii at the high tail serves as one of
many observational tests of the standard $\Lambda$CDM model
\citep[e.g.,][]{2008MNRAS.390.1647B,2012A&A...547A..67W,2012MNRAS.423.2308Z,2020ApJ...898....6A}. 
Among others, \citet{2009MNRAS.392..930O} present detailed predictions
of the largest Einstein radius in the Universe based on the halo model
approach.

Here we compare observed $\theta_{\mathrm{Ein}}$ for the
VENUS clusters with the $\Lambda$CDM prediction of
\citet{2009MNRAS.392..930O}, which adopt a triaxial halo model of
\citet{2002ApJ...574..538J} to derive the distribution of the Einstein
radii,  with some minor updates. The use of a triaxial halo is crucial
given the strong dependence of the lensing strength on the viewing
angle of a cluster. The halo mass function is updated to a fitting
formula presented in \citet{2016MNRAS.456.2486D}. Assuming the
latest best-fit cosmological parameters of \citet{2025JCAP...11..062L}
with $\Omega_{\mathrm{m}}=0.3116$, $\Omega_{\mathrm{b}}=0.04920$,
$\Omega_{\Lambda}=0.6884$, $h=0.6762$, $n_{\mathrm{s}}=0.9709$, and
$\sigma_8=0.8149$, and an effective all-sky area of 40,000~deg$^2$, we
generate a mock halo sample, and for each cluster we assign the
concentration parameter and axis ratios following the probability
distribution function given in \citet{2002ApJ...574..538J}. For each
cluster, the viewing angle is also randomly assigned. We then
compute the tangential critical curve of each cluster to derive the
Einstein radius using the same definition as given in
Equation~\eqref{eq:def_tein}. The Einstein radii are computed for
source redshifts from $z_{\mathrm{s}}=1$ to $9.5$ with an interval of
$\Delta z_{\mathrm{s}}=0.5$, and values of the Einstein radii are
recorded only when they are sufficiently large,
e.g., $\theta_{\mathrm{Ein}}\geq 20''$ for $z_{\mathrm{s}}=1$ and
$\theta_{\mathrm{Ein}}\geq 40''$ for $z_{\mathrm{s}}=9.5$. We repeat this
procedure 800 times to generate 800 realizations of an all-sky
sample of clusters with large Einstein radii.

\begin{center}
    \begin{figure}[!t]
        \centering
    	\includegraphics[scale=0.42]{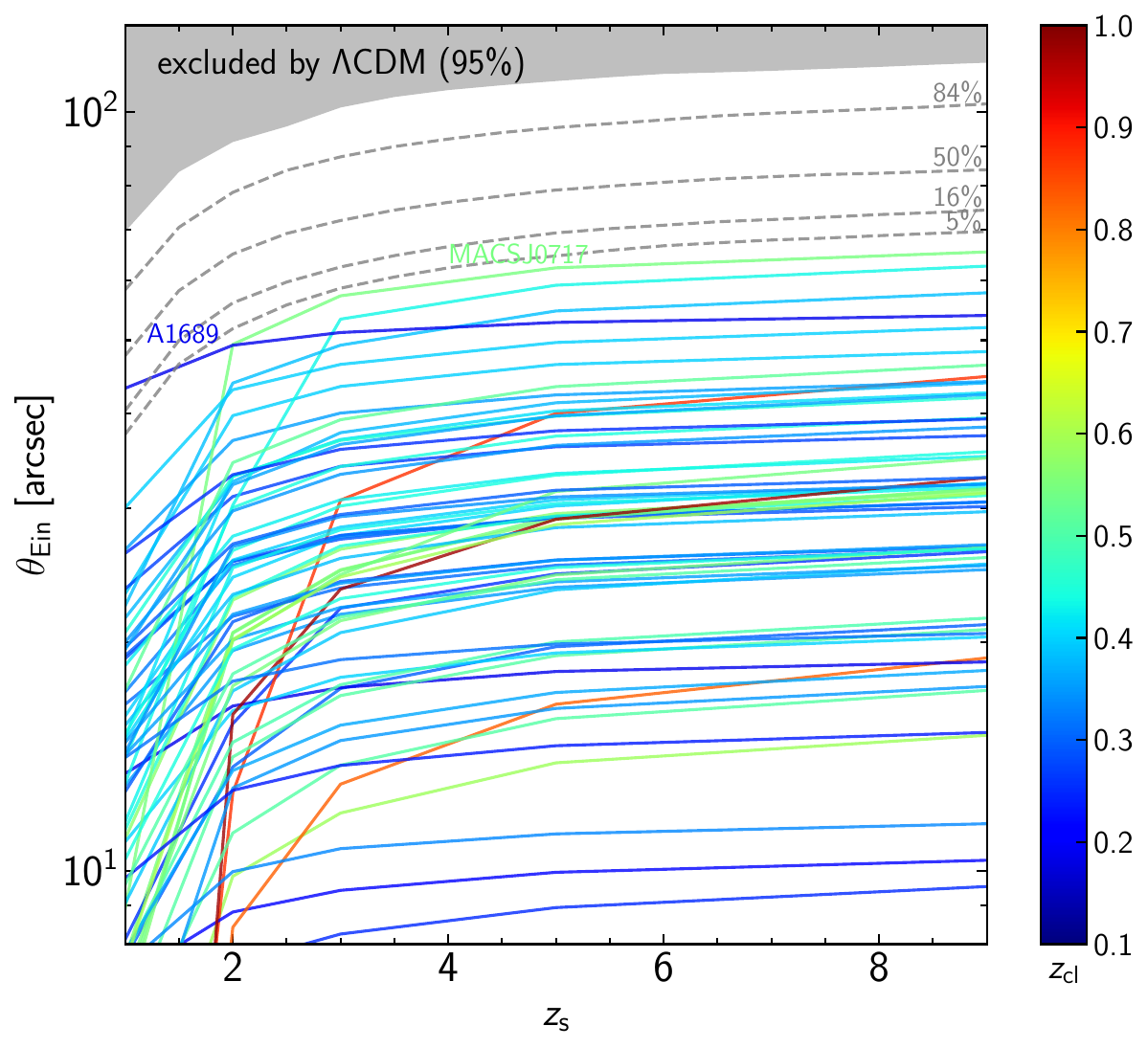}
    	\caption{The Einstein radii $\theta_{\mathrm{Ein}}$ as a
          function of the source redshift $z_{\mathrm{s}}$ for all the
          VENUS clusters are shown by solid lines. The color variation
          is based on the cluster redshift $z_{\mathrm{cl}}$. Dashed
          lines indicate 5\%, 16\%, 50\%, and 84\% limits of the
          largest Einstein radius in the Universe predicted by the
          $\Lambda$CDM model, based on the halo model calculation of
          \citet{2009MNRAS.392..930O} with some minor updates. The
          shaded region is excluded by the $\Lambda$CDM model at the
          95\% confidence limit.}   
        \label{fig:tein_zs}
    \end{figure}
\end{center}

We compare the observed Einstein radii for the VENUS clusters with
the $\Lambda$CDM prediction in Figure~\ref{fig:tein_zs}. The largest
Einstein radius in the VENUS cluster sample is that of Abell~1689 for
$z_{\mathrm{s}}\lesssim 2$ and that of MACS~J0717.5+3745 for
$z_{\mathrm{s}}\gtrsim 2$. We find that the largest Einstein radius in
the VENUS cluster sample is well within the 95\% confidence limit of
the $\Lambda$CDM prediction of the largest Einstein radius in the
Universe, and therefore is compatible with the $\Lambda$CDM model.
It should be noted that the VENUS cluster sample is not complete in
the coverage of clusters with high lensing strengths, and there may be
additional clusters that have even higher Einstein radii, especially
for high $z_{\mathrm{s}}$.

\section{Conclusion}
\label{sec:concl}

We have presented a uniform mass modeling of 60 galaxy clusters
observed with the JWST mostly as part of the VENUS program, using the
{\sc glafic} software. The JWST/NIRCam images are available for 
57 out of the 60 galaxy clusters. We have identified many new multiple 
images, taking advantage of the VENUS JWST/NIRCam
images. Specifically, we have used $\sim 4500$ multiple images for our
mass modeling of the 60 clusters, $\sim 1150$ of which are newly
identified in this paper. As a result, individual VENUS clusters
typically contain $\sim 50$ multiple images.  The large number of
multiple images enables us to tightly constrain the mass distribution of each
cluster. The root-mean-square of differences between observed and
model-predicted multiple image positions is typically $\sim 0\farcs5$,
which is consistent with previous studies. We have conducted the
Jackknife test to show that there is no sign of overfitting for most
of the clusters. The median Einstein radius of the VENUS cluster
sample is $\sim 30''$ and the distribution of the Einstein radii
extends out to $\sim 60''$ for the source redshift of
$z_{\mathrm{s}}=9$, which confirms that lensing strengths of VENUS
clusters are among the highest known to date. For $z_{\mathrm{s}}=9$,
we have identified MACS~J0717.5+3745, MACS~J0600.1$-$2008, and
PLCK~G287.0+32.9 as clusters with the largest Einstein radii
($\theta_{\mathrm{Ein}}\sim 60''$) as well as the largest areas
($\mathcal{A}_{|\mu|>10}\sim 3~\mathrm{arcmin}^2$) of the
high-magnification regions. We have compared the observed Einstein
radii with the $\Lambda$CDM model prediction to conclude that they are
compatible with each other.

The strong lens mass models presented in this paper, together with
independent mass models presented in a companion paper (J.
F. V. Allingham et al., in preparation), will be crucial for studying the
high-redshift Universe with the high-quality VENUS data.  These
regions of highest magnification will enable studies of the faintest
objects at high redshift, helping to push the redshift frontier,
constrain the faint end of the UV luminosity function, and identify
candidate star clusters and metal-poor galaxies for spectroscopic
follow-up.  

\section*{Acknowledgments}

This work was conducted and this paper was written without the use of
AI tools of any kind.
We thank Gabriel Brammer and Hollis Akins for their contributions to
data processing and managements. 
This work is based on observations made with the NASA/ESA/CSA James
Webb Space Telescope. The data were obtained from the Mikulski Archive
for Space Telescopes at the Space Telescope Science Institute, which
is operated by the Association of Universities for Research in
Astronomy, Inc., under NASA contract NAS 5-03127 for JWST. These
observations are associated with program \#6882.
This work was supported by JSPS KAKENHI Grant Numbers JP25H00662 (MO),
JP22K21349 (MO, YF), JP23K13149 (YF), 26K00736 (YF), JP22H04939 (KK),
JP23K20035 (KK), JP24H00004 (KK). 
SF acknowledges support from the Dunlap Institute, funded through an
endowment established by the David Dunlap family and the University of
Toronto. We acknowledge the support of the Canadian Space Agency (CSA)
[25JWGO4A18].
JMD acknowledges the support of projects PID2022-138896NB-C51
(MCIU/AEI/MINECO/FEDER, UE) Ministerio de Ciencia, Investigaci\'on.
MB acknowledges support from the ERC Grant FIRSTLIGHT \# 101053208,
Slovenian national research agency ARIS through grants N1-0238 and
P1-0188, and ESA PRODEX Experiment Arrangements No. 4000146646 and
4000149972.
CJC acknowledges support from the European Research Council (ERC)
Advanced Investigator Grant EPOCHS (788113).
MGO acknowledges financial support from the State Agency for Research
of the Spanish MCIU through Center of Excellence Severo Ochoa award to
the Instituto de Astrof\'{i}sica de Andaluc\'{i}a CEX2021-001131-S
funded by MCIN/AEI/10.13039/501100011033, and from the grant
PID2022-136598NB-C32 “Estallidos8” funded by
MCIN/AEI/10.13039/501100011033 and by “ERDF A way of making
Europe”. MGO also acknowledges the support by the project
ref. AST22\_00001\_Subp\_11 funded from the EU - NextGenerationEU, 
PPCC Junta de Andalucía.
\bibliographystyle{aasjournal}

\bibliography{ref}

\begin{appendix}

\section{References for Multiple Images}
\label{ap:ref}

\subsection{Abell~2744}
Identifications of multiple images from \citet{2011MNRAS.417..333M}, \citet{2014ApJ...797...98L}, \citet{2014MNRAS.444..268R}, \citet{2015MNRAS.452.1437J}, \citet{2014ApJ...786...60A}, \citet{2014ApJ...793L..12Z}, \citet{2018MNRAS.473..663M}, \citet{2023MNRAS.523.4568F}, \citet{2024ApJ...961..186C}, this paper. Spectroscopic redshifts of multiple images from \citet{2018MNRAS.473..663M}, \citet{2023A&A...670A..60B}, \citet{2014ApJ...797...48J}, \citet{2014MNRAS.444..268R}, \citet{2025ApJ...987..186F}, \citet{2025ApJ...982...51P}, \citet{2015ApJ...811...29W}, \citet{2023ApJ...952...84B}, \citet{2024Natur.628...57F}, \citet{2024ApJ...961..186C}.
\subsection{RXC~J0018.5+1626}
Identifications of multiple images from \citet{2011MNRAS.410.1939Z}, \citet{2022MNRAS.516.1373F}, this paper. Spectroscopic redshifts of multiple images from \citet{2022MNRAS.516.1373F}.
\subsection{MACS~J0025.4$-$1222}
Identifications of multiple images from \citet{2008ApJ...687..959B}, \citet{2018ApJ...863..145C}, this paper. Spectroscopic redshifts of multiple images from \citet{2008ApJ...687..959B}.
\subsection{RXC~J0032.1+1808}
Identifications of multiple images from \citet{2020ApJ...898....6A}, this paper, \citet{2017A&A...605A..81D}, \citet{2025ApJ...989..115T}. Spectroscopic redshifts of multiple images from \citet{2017A&A...605A..81D}, \citet{2025ApJ...989..115T}.
\subsection{MACS~J0035.4$-$2015}
Identifications of multiple images from \citet{2023MNRAS.522.5142F}, \citet{2025A&A...698A.161S}, this paper. Spectroscopic redshifts of multiple images from \citet{2025A&A...698A.161S}, \citet{2023MNRAS.522.5142F}.
\subsection{Abell~68}
Identifications of multiple images from \citet{2002MNRAS.330....1S}, \citet{2026ApJ..1001...60C}, \citet{2005MNRAS.359..417S}, \citet{2007ApJ...662..781R}, this paper. Spectroscopic redshifts of multiple images from \citet{2005MNRAS.359..417S}, \citet{2007ApJ...662..781R}.
\subsection{Abell~2813}
Identifications of multiple images from this paper. 
\subsection{El~Gordo}
Identifications of multiple images from \citet{2013ApJ...770L..15Z}, \citet{2023A&A...678A...3C}, \citet{2021ApJ...908..146C}, \citet{2020ApJ...904..106D}, \citet{2023A&A...672A...3D}, \citet{2023ApJ...952...81F}, this paper. Spectroscopic redshifts of multiple images from \citet{2023A&A...678A...3C}, \citet{2021ApJ...908..146C}, \citet{2023ApJ...952...81F}.
\subsection{Abell~209}
Identifications of multiple images from \citet{2015ApJ...801...44Z}. Spectroscopic redshifts of multiple images from \citet{2019A&A...632A..36C}.
\subsection{WL~J013719$-$08284}
Identifications of multiple images from \citet{2022Natur.603..815W}, \citet{2023ApJ...945...53V}, Welch~et~al.,~in~prep., this paper. Spectroscopic redshifts of multiple images from \citet{2025ApJ...988L..76P}, Welch~et~al.,~in~prep..
\subsection{CL~J0152.7$-$1357}
Identifications of multiple images from \citet{2019ApJ...874..132A}, this paper. Spectroscopic redshifts of multiple images from \citet{2019ApJ...874..132A}.
\subsection{MACS~J0159.8$-$0849}
Identifications of multiple images from \citet{2018ApJ...863..145C}, this paper. 
\subsection{RXC~J0232.2$-$4420}
Identifications of multiple images from \citet{2026ApJ..1001...60C}, this paper. Spectroscopic redshifts of multiple images from \citet{2019PhDT.......198M}.
\subsection{Abell~370}
Identifications of multiple images from \citet{2010MNRAS.402L..44R}, \citet{2018MNRAS.473.4279D}, \citet{2025ApJ...987..186F}, Fudamoto~et~al.,~in~prep., \citet{2014MNRAS.444..268R}, \citet{2018ApJ...855....4K}, \citet{2017MNRAS.469.3946L}, \citet{2022Natur.611..256C}, \citet{2019MNRAS.485.3738L}, this paper. Spectroscopic redshifts of multiple images from \citet{1988A&A...191L..19S}, \citet{2010MNRAS.402L..44R}, \citet{2018MNRAS.473.4279D}, \citet{2017MNRAS.469.3946L}, \citet{2024ApJ...973...77G}, \citet{2021MNRAS.506.6144G}, \citet{2019MNRAS.485.3738L}, \citet{2026ApJS..282....3S}, \citet{2021A&A...646A..83R}, \citet{2025ApJ...987..186F}.
\subsection{Abell~S295}
Identifications of multiple images from \citet{2018ApJ...863..145C}, this paper. Spectroscopic redshifts of multiple images from \citet{1994A&A...289L..34E}.
\subsection{Abell~383}
Identifications of multiple images from \citet{2025MNRAS.541.2341C}, \citet{2011ApJ...742..117Z}, \citet{2001ApJ...552..493S}, \citet{2011MNRAS.414L..31R}. Spectroscopic redshifts of multiple images from \citet{2025MNRAS.541.2341C}, \citet{2001ApJ...552..493S}, \citet{2011ApJ...728L..39N}, \citet{2011MNRAS.414L..31R}, \citet{2015ApJ...801...44Z}.
\subsection{MACS~J0257.1$-$2325}
Identifications of multiple images from \citet{2011MNRAS.410.1939Z}, this paper, \citet{2026ApJ..1010..237N}. Spectroscopic redshifts of multiple images from \citet{2026ApJ..1010..237N}.
\subsection{MACS~J0257.6$-$2209}
Identifications of multiple images from \citet{2021A&A...646A..83R}, this paper. Spectroscopic redshifts of multiple images from \citet{2021A&A...646A..83R}.
\subsection{MACS~J0308.9+2645}
Identifications of multiple images from \citet{2018ApJ...858...42A}, this paper, \citet{2025arXiv251208054A}. 
\subsection{PLCK~G171.9$-$40.7}
Identifications of multiple images from \citet{2018ApJ...858...42A}, this paper. 
\subsection{MACS~J0329.7$-$0211}
Identifications of multiple images from \citet{2019A&A...632A..36C}, \citet{2015ApJ...801...44Z}, \citet{2021A&A...646A..83R}, this paper. Spectroscopic redshifts of multiple images from \citet{2019A&A...632A..36C}, \citet{2015ApJ...801...44Z}, \citet{2021A&A...646A..83R}.
\subsection{Abell~3192}
Identifications of multiple images from \citet{2013MNRAS.429..833H}, \citet{2012ApJ...748L..23H}, \citet{2020MNRAS.496.2591O}, \citet{2026ApJ..1001...60C}, this paper. Spectroscopic redshifts of multiple images from \citet{2013MNRAS.429..833H}.
\subsection{MACS~J0416.1$-$2403}
Identifications of multiple images from \citet{2013ApJ...762L..30Z}, \citet{2014MNRAS.443.1549J}, \citet{2016ApJ...819..114K}, \citet{2018ApJ...855....4K}, \citet{2015MNRAS.447.3130D}, \citet{2017A&A...600A..90C}, \citet{2021A&A...646A..83R}, \citet{2023A&A...674A..79B}, \citet{2024A&A...690A.114D}, this paper. Spectroscopic redshifts of multiple images from \citet{2015ApJ...800...38G}, \citet{2012MNRAS.427.1953C}, \citet{2016ApJ...831..182H}, \citet{2025A&A...696A..15R}, \citet{2018NatAs...2..324R}, \citet{2023A&A...674A..79B}, \citet{2017A&A...600A..90C}, \citet{2021A&A...646A..83R}, \citet{2026ApJS..282....3S}.
\subsection{MACS~J0417.5$-$1154}
Identifications of multiple images from \citet{2019ApJ...873...96M}, this paper, \citet{2025ApJ...989..115T}. Spectroscopic redshifts of multiple images from \citet{2019ApJ...873...96M}, \citet{2026ApJS..282....3S}, \citet{2019MNRAS.483.3082J}, \citet{2025ApJ...989..115T}.
\subsection{MACS~J0429.6$-$0253}
Identifications of multiple images from \citet{2019A&A...632A..36C}, \citet{2015ApJ...801...44Z}, this paper. Spectroscopic redshifts of multiple images from \citet{2019A&A...632A..36C}.
\subsection{MACS~J0451.9+0006}
Identifications of multiple images from MUSE~Atlas, \citet{2011MNRAS.413..643R}, this paper. Spectroscopic redshifts of multiple images from MUSE~Atlas, \citet{2011MNRAS.413..643R}.
\subsection{Abell~521}
Identifications of multiple images from \citet{2010MNRAS.404..325R}, \citet{2022MNRAS.516.2420M}, this paper. Spectroscopic redshifts of multiple images from \citet{2010MNRAS.404..325R}.
\subsection{MS~0451.6$-$0305}
Identifications of multiple images from \citet{2004MNRAS.352..759B}, \citet{2021MNRAS.508.1206J}, \citet{2014MNRAS.445..201M}, this paper. Spectroscopic redshifts of multiple images from \citet{2004MNRAS.352..759B}, \citet{2021MNRAS.508.1206J}, \citet{2016MNRAS.462L...6K}.
\subsection{MACS~J0553.4$-$3342}
Identifications of multiple images from \citet{2017MNRAS.471.3305E}, \citet{2026ApJ..1001...60C}, this paper. Spectroscopic redshifts of multiple images from \citet{2017MNRAS.471.3305E}, \citet{2026ApJ..1001...60C}.
\subsection{MACS~J0600.1$-$2008}
Identifications of multiple images from \citet{2024MNRAS.533.2242F}, \citet{2021ApJ...911...99F}, this paper. Spectroscopic redshifts of multiple images from \citet{2024MNRAS.533.2242F}, \citet{2021ApJ...911...99F}, \citet{2021MNRAS.505.4838L}.
\subsection{S~J0615$-$5746}
Identifications of multiple images from \citet{2018ApJ...863..154P}, \citet{2024Natur.632..513A}, \citet{2026A&A...705A.173M}, this paper. Spectroscopic redshifts of multiple images from \citet{2018ApJ...863..154P}, \citet{2026A&A...705A.173M}.
\subsection{Bullet~cluster}
Identifications of multiple images from \citet{2001A&A...379...96M}, \citet{2009ApJ...706.1201B}, \citet{2016A&A...594A.121P}, \citet{2006ApJ...652..937B}, \citet{2021A&A...646A..83R}, \citet{2025ApJ...987L..15C}, this paper, \citet{2026A&A...710A.207R}. Spectroscopic redshifts of multiple images from \citet{2001A&A...379...96M}, \citet{2026A&A...710A.207R}, \citet{2021A&A...646A..83R}, \citet{2010ApJ...720..245G}, \citet{2025ApJ...987L..15C}, \citet{2016A&A...594A.121P}.
\subsection{MACS~J0717.5+3745}
Identifications of multiple images from \citet{2009ApJ...707L.102Z}, \citet{2012A&A...544A..71L}, \citet{2014ApJ...783L..12V}, \citet{2015MNRAS.451.3920D}, \citet{2016ApJ...819..114K}, this paper. Spectroscopic redshifts of multiple images from \citet{2012A&A...544A..71L}, \citet{2014ApJ...782L..36S}, \citet{2015ApJ...812..114T}, \citet{2014ApJ...783L..12V}.
\subsection{S~J0723.3$-$7327}
Identifications of multiple images from \citet{2022ApJ...938...14G}, \citet{2023ApJ...945...49M}, \citet{2022ApJ...938L...6P}, this paper. Spectroscopic redshifts of multiple images from \citet{2022ApJ...938...14G}, \citet{2023ApJ...945...49M}, \citet{2023MNRAS.525.1867N}.
\subsection{RXC~J0911.1+1746}
Identifications of multiple images from this paper, \citet{2020ApJ...888..124S}. 
\subsection{MACS~J0940.9+0744}
Identifications of multiple images from \citet{2019MNRAS.489.5022C}, \citet{2021A&A...646A..83R}, this paper. Spectroscopic redshifts of multiple images from \citet{2019MNRAS.489.5022C}, \citet{2021A&A...646A..83R}.
\subsection{RXC~J0949.8+1707}
Identifications of multiple images from \citet{2023MNRAS.522.1118A}. Spectroscopic redshifts of multiple images from \citet{2023MNRAS.522.1118A}.
\subsection{MACS~J1115.8+0129}
Identifications of multiple images from \citet{2015ApJ...801...44Z}, \citet{2019A&A...632A..36C}, this paper. Spectroscopic redshifts of multiple images from \citet{2019A&A...632A..36C}.
\subsection{PLCK~G165.7+67.0}
Identifications of multiple images from \citet{2024ApJ...961..171F}, \citet{2015A&A...581A.105C}, \citet{2018A&A...620A..60C}, \citet{2019ApJ...871...51F}. Spectroscopic redshifts of multiple images from \citet{2023A&A...675L...4P}, \citet{2015A&A...581A.105C}, \citet{2024ApJ...961..171F}.
\subsection{MACS~J1149.5+2223}
Identifications of multiple images from \citet{2009ApJ...703L.132Z}, \citet{2015Sci...347.1123K}, \citet{2023ApJ...948...93K}, \citet{2009ApJ...707L.163S}, \citet{2014MNRAS.443..957R}, \citet{2012Natur.489..406Z}, \citet{2016MNRAS.457.2029J}, \citet{2016ApJ...817...60T}, \citet{2016ApJ...819..114K}, \citet{2014MNRAS.444..268R}, \citet{2024A&A...689A..42S}, this paper. Spectroscopic redshifts of multiple images from \citet{2009ApJ...707L.163S}, \citet{2015Sci...347.1123K}, \citet{2016MNRAS.457.2029J}, \citet{2016ApJ...822...78G}, \citet{2016ApJ...817...60T}, \citet{2026ApJS..282....3S}, \citet{2024A&A...689A..42S}.
\subsection{PLCK~G287.0+32.9}
Identifications of multiple images from \citet{2017ApJ...839L..11Z}, \citet{2024A&A...686A...4D}, this paper, \citet{2026arXiv260931944B}. Spectroscopic redshifts of multiple images from \citet{2024A&A...686A...4D}.
\subsection{MACS~J1206.2$-$0847}
Identifications of multiple images from \citet{2017A&A...607A..93C}, \citet{2009MNRAS.395.1213E}, \citet{2012ApJ...749...97Z}, \citet{2021A&A...646A..83R}, this paper. Spectroscopic redshifts of multiple images from \citet{2017A&A...607A..93C}, \citet{2009MNRAS.395.1213E}, \citet{2012ApJ...749...97Z}, \citet{2021A&A...646A..83R}.
\subsection{MACS~J1311.0$-$0310}
Identifications of multiple images from \citet{2015ApJ...801...44Z}, this paper. Spectroscopic redshifts of multiple images from \citet{2019A&A...632A..36C}.
\subsection{Abell~1689}
Identifications of multiple images from \citet{2005ApJ...621...53B}, \citet{2007ApJ...668..643L}, \citet{2010ApJ...723.1678C}, \citet{2016A&A...590A..14B}, \citet{2015MNRAS.446..683D}, this paper. Spectroscopic redshifts of multiple images from \citet{2005ApJ...621...53B}, \citet{2007ApJ...665..921F}, \citet{2010Sci...329..924J}, \citet{2002ApJ...568..558F}, \citet{2016A&A...590A..14B}, \citet{2007ApJ...668..643L}.
\subsection{RXC~J1347.5$-$1145}
Identifications of multiple images from \citet{2008A&A...481...65H}, \citet{2019A&A...632A..36C}, \citet{2005A&A...437...49B}, \citet{2015ApJ...801...44Z}, \citet{2021A&A...646A..83R}. Spectroscopic redshifts of multiple images from \citet{2002ApJ...577..133R}, \citet{2019A&A...632A..36C}, \citet{2002ApJ...573..524C}, \citet{2021A&A...646A..83R}.
\subsection{Abell~1835}
Identifications of multiple images from \citet{2010MNRAS.404..325R}, \citet{2012MNRAS.425.2069M}. Spectroscopic redshifts of multiple images from \citet{2010MNRAS.404..325R}.
\subsection{MACS~J1423.8+2404}
Identifications of multiple images from \citet{2010MNRAS.405..777L}, \citet{2015ApJ...801...44Z}, this paper. Spectroscopic redshifts of multiple images from \citet{2010MNRAS.405..777L}, \citet{2026ApJS..282....3S}.
\subsection{RXC~J1532.9+3021}
Identifications of multiple images from \citet{2015ApJ...801...44Z}, this paper. 
\subsection{PLCK~G004.5$-$19.5}
Identifications of multiple images from \citet{2014A&A...562A..43S}, this paper, \citet{2026arXiv260106015Y}. Spectroscopic redshifts of multiple images from \citet{2014A&A...562A..43S}.
\subsection{MACS~J1931.8$-$2635}
Identifications of multiple images from \citet{2019A&A...632A..36C}, \citet{2015ApJ...801...44Z}, this paper, \citet{2026ApJ..1005..215A}, \citet{2026arXiv260104156C}. Spectroscopic redshifts of multiple images from \citet{2019A&A...632A..36C}, \citet{2015ApJ...801...44Z}, \citet{2026arXiv260104156C}.
\subsection{S~J2031.8$-$4036}
Identifications of multiple images from \citet{2015MNRAS.446L..16R}, \citet{2021A&A...646A..83R}, \citet{2026ApJ..1001...60C}, this paper. Spectroscopic redshifts of multiple images from \citet{2012MNRAS.427.1953C}, \citet{2015MNRAS.446L..16R}, \citet{2021A&A...646A..83R}.
\subsection{MACS~J2129.4$-$0741}
Identifications of multiple images from \citet{2019A&A...632A..36C}, \citet{2015ApJ...801...44Z}, \citet{2017MNRAS.466.4094M}, \citet{2021MNRAS.508.1206J}, \citet{2026ApJ..1001...60C}, this paper. Spectroscopic redshifts of multiple images from \citet{2015ApJ...812..114T}, \citet{2019A&A...632A..36C}, \citet{2012MNRAS.427.1953C}, \citet{2017MNRAS.466.4094M}, \citet{2016ApJ...823L..14H}, \citet{2021MNRAS.508.1206J}.
\subsection{RXC~J2129.7+0005}
Identifications of multiple images from \citet{2019A&A...632A..36C}, \citet{2015ApJ...801...44Z}, \citet{2010MNRAS.404..325R}, \citet{2023Sci...380..416W}, \citet{2026ApJ..1008..230D}, this paper. Spectroscopic redshifts of multiple images from \citet{2019A&A...632A..36C}, \citet{2013ApJ...772..141B}, \citet{2023Sci...380..416W}.
\subsection{S~J2131.1$-$4019}
Identifications of multiple images from \citet{2021A&A...646A..83R}, this paper. Spectroscopic redshifts of multiple images from \citet{2021A&A...646A..83R}.
\subsection{MS~2137$-$2353}
Identifications of multiple images from \citet{1993ApJ...407...33M}, \citet{2015ApJ...801...44Z}, \citet{2025MNRAS.541.2341C}, this paper. Spectroscopic redshifts of multiple images from \citet{2002ApJ...574L.129S}, \citet{2015ApJ...801...44Z}, \citet{2025MNRAS.541.2341C}.
\subsection{RXC~J2211.7$-$0350}
Identifications of multiple images from \citet{2018ApJ...859..159C}, \citet{2026ApJ..1001...60C}, \citet{2025arXiv251205180Z}. Spectroscopic redshifts of multiple images from \citet{2018ApJ...859..159C}, \citet{2025arXiv251205180Z}.
\subsection{MACS~J2214.9$-$1359}
Identifications of multiple images from \citet{2011MNRAS.410.1939Z}, \citet{2021A&A...646A..83R}, this paper. Spectroscopic redshifts of multiple images from \citet{2021A&A...646A..83R}.
\subsection{Abell~S1063}
Identifications of multiple images from \citet{2014MNRAS.444..268R}, \citet{2014ApJ...797...48J}, \citet{2016A&A...587A..80C}, \citet{2016MNRAS.459.3447D}, \citet{2018ApJ...855....4K}, \citet{2017A&A...599A..28K}, \citet{2024MNRAS.527.3246B}, \citet{2015ApJ...801...44Z}, \citet{2026arXiv260111704D}. Spectroscopic redshifts of multiple images from \citet{2017A&A...599A..28K}, \citet{2015A&A...574A..11K}, \citet{2016A&A...587A..80C}, \citet{2013A&A...559L...9B}, \citet{2015ApJ...812..114T}, \citet{2024MNRAS.527.3246B}.
\subsection{Abell~2537}
Identifications of multiple images from \citet{2018ApJ...859..159C}, this paper. Spectroscopic redshifts of multiple images from \citet{2018ApJ...859..159C}.
\subsection{SPT$-$CL~J2325$-$4111}
Identifications of multiple images from \citet{2025ApJ...980...48M}, this paper. Spectroscopic redshifts of multiple images from \citet{2025ApJ...980...48M}.

\end{appendix}

\end{document}